%% file: main.tex
\pdfoutput=1
\documentclass[preptrint, 12pt]{elsarticle}

\usepackage{natbib}
    \usepackage{graphicx}
    \usepackage[table,xcdraw]{xcolor}
    \usepackage{tikz}
    \usepackage{makecell}
    \usepackage{dsfont}
    \usepackage{multirow}
    \usepackage{verbatim}
    \usetikzlibrary{mindmap,shadows}

\usepackage{tikz}
\usepackage{amsmath}
\usepackage{graphicx}

    \usepackage[ruled,vlined]{algorithm2e}

    \usepackage{amsmath}

    \usepackage{graphicx}

    \usepackage{svg}

    \usepackage{caption}

    \usepackage[Export]{adjustbox} 
    \adjustboxset{max size={0.9\linewidth}{0.9\paperheight}}
    \usepackage{float}
    \usepackage{xcolor} 
    \usepackage{adjustbox}
    \usepackage{pdfpages}

    \usepackage{enumerate} 
    \usepackage{geometry} 
    \usepackage{amsmath} 
    \usepackage{amssymb} 
    \usepackage{textcomp} 
    \AtBeginDocument{%
    }
    \usepackage{upquote} 
    \usepackage{eurosym} 
    \usepackage[mathletters]{ucs} 
    \usepackage{fancyvrb} 
    \usepackage{bera}
    \usepackage{grffile} 
    \makeatletter 
    \def\Gread@@xetex#1{%
      \IfFileExists{"\Gin@base".bb}%
      {\Gread@eps{\Gin@base.bb}}%
      {\Gread@@xetex@aux#1}%
    }
    \makeatother

    \usepackage{longtable} 
    \usepackage{booktabs}  
    \usepackage[inline]{enumitem} 
    \usepackage[normalem]{ulem} 
    \usepackage{mathrsfs}
    
    \usepackage[perpage]{footmisc} 

    \usepackage{hyperref}
    \hypersetup{colorlinks,allcolors=black}

    \definecolor{urlcolor}{rgb}{0,.145,.698}
    \definecolor{linkcolor}{rgb}{.71,0.21,0.01}
    \definecolor{citecolor}{rgb}{.12,.54,.11}

    \definecolor{ansi-black}{HTML}{3E424D}
    \definecolor{ansi-black-intense}{HTML}{282C36}
    \definecolor{ansi-red}{HTML}{E75C58}
    \definecolor{ansi-red-intense}{HTML}{B22B31}
    \definecolor{ansi-green}{HTML}{00A250}
    \definecolor{ansi-green-intense}{HTML}{007427}
    \definecolor{ansi-yellow}{HTML}{DDB62B}
    \definecolor{ansi-yellow-intense}{HTML}{B27D12}
    \definecolor{ansi-blue}{HTML}{208FFB}
    \definecolor{ansi-blue-intense}{HTML}{0065CA}
    \definecolor{ansi-magenta}{HTML}{D160C4}
    \definecolor{ansi-magenta-intense}{HTML}{A03196}
    \definecolor{ansi-cyan}{HTML}{60C6C8}
    \definecolor{ansi-cyan-intense}{HTML}{258F8F}
    \definecolor{ansi-white}{HTML}{C5C1B4}
    \definecolor{ansi-white-intense}{HTML}{A1A6B2}
    \definecolor{ansi-default-inverse-fg}{HTML}{FFFFFF}
    \definecolor{ansi-default-inverse-bg}{HTML}{000000}

    \DefineVerbatimEnvironment{Highlighting}{Verbatim}{commandchars=\\\{\}}

    \newcommand{\GreenCircle}[1][blue!40!green,fill=blue!40!green]{\tikz[baseline=-0.5ex]\draw[#1,radius=3pt] (0,0) circle ;}%
    \newcommand{\OrangeCircle}[1][red!30!yellow,fill=red!30!yellow]{\tikz[baseline=-0.5ex]\draw[#1,radius=3pt] (0,0) circle ;}%
    \newcommand{\RedCircle}[1][black!10!red,fill=black!10!red]{\tikz[baseline=-0.5ex]\draw[#1,radius=3pt] (0,0) circle ;}%
    \newcommand{\GrayCircle}[1][gray,fill=gray]{\tikz[baseline=-0.5ex]\draw[#1,radius=3pt] (0,0) circle ;}%
    
    \let\Oldtex\TeX
    \let\Oldlatex\LaTeX
    \renewcommand{\TeX}{\textrm{\Oldtex}}
    \renewcommand{\LaTeX}{\textrm{\Oldlatex}}

\newcommand{\Hquad}{\hspace{0.5em}}

\makeatletter
\def\PY@reset{\let\PY@it=\relax \let\PY@bf=\relax%
    \let\PY@ul=\relax \let\PY@tc=\relax%
    \let\PY@bc=\relax \let\PY@ff=\relax}
\def\PY@tok#1{\csname PY@tok@#1\endcsname}
\def\PY@toks#1+{\ifx\relax#1\empty\else%
    \PY@tok{#1}\expandafter\PY@toks\fi}
\def\PY@do#1{\PY@bc{\PY@tc{\PY@ul{%
    \PY@it{\PY@bf{\PY@ff{#1}}}}}}}
\def\PY#1#2{\PY@reset\PY@toks#1+\relax+\PY@do{#2}}

\expandafter\def\csname PY@tok@w\endcsname{\def\PY@tc##1{\textcolor[rgb]{0.73,0.73,0.73}{##1}}}
\expandafter\def\csname PY@tok@c\endcsname{\let\PY@it=\textit\def\PY@tc##1{\textcolor[rgb]{0.25,0.50,0.50}{##1}}}
\expandafter\def\csname PY@tok@cp\endcsname{\def\PY@tc##1{\textcolor[rgb]{0.74,0.48,0.00}{##1}}}
\expandafter\def\csname PY@tok@k\endcsname{\let\PY@bf=\textbf\def\PY@tc##1{\textcolor[rgb]{0.00,0.50,0.00}{##1}}}
\expandafter\def\csname PY@tok@kp\endcsname{\def\PY@tc##1{\textcolor[rgb]{0.00,0.50,0.00}{##1}}}
\expandafter\def\csname PY@tok@kt\endcsname{\def\PY@tc##1{\textcolor[rgb]{0.69,0.00,0.25}{##1}}}
\expandafter\def\csname PY@tok@o\endcsname{\def\PY@tc##1{\textcolor[rgb]{0.40,0.40,0.40}{##1}}}
\expandafter\def\csname PY@tok@ow\endcsname{\let\PY@bf=\textbf\def\PY@tc##1{\textcolor[rgb]{0.67,0.13,1.00}{##1}}}
\expandafter\def\csname PY@tok@nb\endcsname{\def\PY@tc##1{\textcolor[rgb]{0.00,0.50,0.00}{##1}}}
\expandafter\def\csname PY@tok@nf\endcsname{\def\PY@tc##1{\textcolor[rgb]{0.00,0.00,1.00}{##1}}}
\expandafter\def\csname PY@tok@nc\endcsname{\let\PY@bf=\textbf\def\PY@tc##1{\textcolor[rgb]{0.00,0.00,1.00}{##1}}}
\expandafter\def\csname PY@tok@nn\endcsname{\let\PY@bf=\textbf\def\PY@tc##1{\textcolor[rgb]{0.00,0.00,1.00}{##1}}}
\expandafter\def\csname PY@tok@ne\endcsname{\let\PY@bf=\textbf\def\PY@tc##1{\textcolor[rgb]{0.82,0.25,0.23}{##1}}}
\expandafter\def\csname PY@tok@nv\endcsname{\def\PY@tc##1{\textcolor[rgb]{0.10,0.09,0.49}{##1}}}
\expandafter\def\csname PY@tok@no\endcsname{\def\PY@tc##1{\textcolor[rgb]{0.53,0.00,0.00}{##1}}}
\expandafter\def\csname PY@tok@nl\endcsname{\def\PY@tc##1{\textcolor[rgb]{0.63,0.63,0.00}{##1}}}
\expandafter\def\csname PY@tok@ni\endcsname{\let\PY@bf=\textbf\def\PY@tc##1{\textcolor[rgb]{0.60,0.60,0.60}{##1}}}
\expandafter\def\csname PY@tok@na\endcsname{\def\PY@tc##1{\textcolor[rgb]{0.49,0.56,0.16}{##1}}}
\expandafter\def\csname PY@tok@nt\endcsname{\let\PY@bf=\textbf\def\PY@tc##1{\textcolor[rgb]{0.00,0.50,0.00}{##1}}}
\expandafter\def\csname PY@tok@nd\endcsname{\def\PY@tc##1{\textcolor[rgb]{0.67,0.13,1.00}{##1}}}
\expandafter\def\csname PY@tok@s\endcsname{\def\PY@tc##1{\textcolor[rgb]{0.73,0.13,0.13}{##1}}}
\expandafter\def\csname PY@tok@sd\endcsname{\let\PY@it=\textit\def\PY@tc##1{\textcolor[rgb]{0.73,0.13,0.13}{##1}}}
\expandafter\def\csname PY@tok@si\endcsname{\let\PY@bf=\textbf\def\PY@tc##1{\textcolor[rgb]{0.73,0.40,0.53}{##1}}}
\expandafter\def\csname PY@tok@se\endcsname{\let\PY@bf=\textbf\def\PY@tc##1{\textcolor[rgb]{0.73,0.40,0.13}{##1}}}
\expandafter\def\csname PY@tok@sr\endcsname{\def\PY@tc##1{\textcolor[rgb]{0.73,0.40,0.53}{##1}}}
\expandafter\def\csname PY@tok@ss\endcsname{\def\PY@tc##1{\textcolor[rgb]{0.10,0.09,0.49}{##1}}}
\expandafter\def\csname PY@tok@sx\endcsname{\def\PY@tc##1{\textcolor[rgb]{0.00,0.50,0.00}{##1}}}
\expandafter\def\csname PY@tok@m\endcsname{\def\PY@tc##1{\textcolor[rgb]{0.40,0.40,0.40}{##1}}}
\expandafter\def\csname PY@tok@gh\endcsname{\let\PY@bf=\textbf\def\PY@tc##1{\textcolor[rgb]{0.00,0.00,0.50}{##1}}}
\expandafter\def\csname PY@tok@gu\endcsname{\let\PY@bf=\textbf\def\PY@tc##1{\textcolor[rgb]{0.50,0.00,0.50}{##1}}}
\expandafter\def\csname PY@tok@gd\endcsname{\def\PY@tc##1{\textcolor[rgb]{0.63,0.00,0.00}{##1}}}
\expandafter\def\csname PY@tok@gi\endcsname{\def\PY@tc##1{\textcolor[rgb]{0.00,0.63,0.00}{##1}}}
\expandafter\def\csname PY@tok@gr\endcsname{\def\PY@tc##1{\textcolor[rgb]{1.00,0.00,0.00}{##1}}}
\expandafter\def\csname PY@tok@ge\endcsname{\let\PY@it=\textit}
\expandafter\def\csname PY@tok@gs\endcsname{\let\PY@bf=\textbf}
\expandafter\def\csname PY@tok@gp\endcsname{\let\PY@bf=\textbf\def\PY@tc##1{\textcolor[rgb]{0.00,0.00,0.50}{##1}}}
\expandafter\def\csname PY@tok@go\endcsname{\def\PY@tc##1{\textcolor[rgb]{0.53,0.53,0.53}{##1}}}
\expandafter\def\csname PY@tok@gt\endcsname{\def\PY@tc##1{\textcolor[rgb]{0.00,0.27,0.87}{##1}}}
\expandafter\def\csname PY@tok@err\endcsname{\def\PY@bc##1{\setlength{\fboxsep}{0pt}\fcolorbox[rgb]{1.00,0.00,0.00}{1,1,1}{\strut ##1}}}
\expandafter\def\csname PY@tok@kc\endcsname{\let\PY@bf=\textbf\def\PY@tc##1{\textcolor[rgb]{0.00,0.50,0.00}{##1}}}
\expandafter\def\csname PY@tok@kd\endcsname{\let\PY@bf=\textbf\def\PY@tc##1{\textcolor[rgb]{0.00,0.50,0.00}{##1}}}
\expandafter\def\csname PY@tok@kn\endcsname{\let\PY@bf=\textbf\def\PY@tc##1{\textcolor[rgb]{0.00,0.50,0.00}{##1}}}
\expandafter\def\csname PY@tok@kr\endcsname{\let\PY@bf=\textbf\def\PY@tc##1{\textcolor[rgb]{0.00,0.50,0.00}{##1}}}
\expandafter\def\csname PY@tok@bp\endcsname{\def\PY@tc##1{\textcolor[rgb]{0.00,0.50,0.00}{##1}}}
\expandafter\def\csname PY@tok@fm\endcsname{\def\PY@tc##1{\textcolor[rgb]{0.00,0.00,1.00}{##1}}}
\expandafter\def\csname PY@tok@vc\endcsname{\def\PY@tc##1{\textcolor[rgb]{0.10,0.09,0.49}{##1}}}
\expandafter\def\csname PY@tok@vg\endcsname{\def\PY@tc##1{\textcolor[rgb]{0.10,0.09,0.49}{##1}}}
\expandafter\def\csname PY@tok@vi\endcsname{\def\PY@tc##1{\textcolor[rgb]{0.10,0.09,0.49}{##1}}}
\expandafter\def\csname PY@tok@vm\endcsname{\def\PY@tc##1{\textcolor[rgb]{0.10,0.09,0.49}{##1}}}
\expandafter\def\csname PY@tok@sa\endcsname{\def\PY@tc##1{\textcolor[rgb]{0.73,0.13,0.13}{##1}}}
\expandafter\def\csname PY@tok@sb\endcsname{\def\PY@tc##1{\textcolor[rgb]{0.73,0.13,0.13}{##1}}}
\expandafter\def\csname PY@tok@sc\endcsname{\def\PY@tc##1{\textcolor[rgb]{0.73,0.13,0.13}{##1}}}
\expandafter\def\csname PY@tok@dl\endcsname{\def\PY@tc##1{\textcolor[rgb]{0.73,0.13,0.13}{##1}}}
\expandafter\def\csname PY@tok@s2\endcsname{\def\PY@tc##1{\textcolor[rgb]{0.73,0.13,0.13}{##1}}}
\expandafter\def\csname PY@tok@sh\endcsname{\def\PY@tc##1{\textcolor[rgb]{0.73,0.13,0.13}{##1}}}
\expandafter\def\csname PY@tok@s1\endcsname{\def\PY@tc##1{\textcolor[rgb]{0.73,0.13,0.13}{##1}}}
\expandafter\def\csname PY@tok@mb\endcsname{\def\PY@tc##1{\textcolor[rgb]{0.40,0.40,0.40}{##1}}}
\expandafter\def\csname PY@tok@mf\endcsname{\def\PY@tc##1{\textcolor[rgb]{0.40,0.40,0.40}{##1}}}
\expandafter\def\csname PY@tok@mh\endcsname{\def\PY@tc##1{\textcolor[rgb]{0.40,0.40,0.40}{##1}}}
\expandafter\def\csname PY@tok@mi\endcsname{\def\PY@tc##1{\textcolor[rgb]{0.40,0.40,0.40}{##1}}}
\expandafter\def\csname PY@tok@il\endcsname{\def\PY@tc##1{\textcolor[rgb]{0.40,0.40,0.40}{##1}}}
\expandafter\def\csname PY@tok@mo\endcsname{\def\PY@tc##1{\textcolor[rgb]{0.40,0.40,0.40}{##1}}}
\expandafter\def\csname PY@tok@ch\endcsname{\let\PY@it=\textit\def\PY@tc##1{\textcolor[rgb]{0.25,0.50,0.50}{##1}}}
\expandafter\def\csname PY@tok@cm\endcsname{\let\PY@it=\textit\def\PY@tc##1{\textcolor[rgb]{0.25,0.50,0.50}{##1}}}
\expandafter\def\csname PY@tok@cpf\endcsname{\let\PY@it=\textit\def\PY@tc##1{\textcolor[rgb]{0.25,0.50,0.50}{##1}}}
\expandafter\def\csname PY@tok@c1\endcsname{\let\PY@it=\textit\def\PY@tc##1{\textcolor[rgb]{0.25,0.50,0.50}{##1}}}
\expandafter\def\csname PY@tok@cs\endcsname{\let\PY@it=\textit\def\PY@tc##1{\textcolor[rgb]{0.25,0.50,0.50}{##1}}}

\makeatother

    \makeatletter
        \newbox\Wrappedcontinuationbox 
        \newbox\Wrappedvisiblespacebox 
        \newcommand*\Wrappedvisiblespace {\textcolor{red}{\textvisiblespace}} 
        \newcommand*\Wrappedcontinuationsymbol {\textcolor{red}{\llap{\tiny$\m@th\hookrightarrow$}}} 
        \newcommand*\Wrappedcontinuationindent {3ex } 
        \newcommand*\Wrappedafterbreak {\kern\Wrappedcontinuationindent\copy\Wrappedcontinuationbox} 
        \newcommand*\Wrappedbreaksatspecials {%
            \def\PYGZus{\discretionary{\char`\_}{\Wrappedafterbreak}{\char`\_}}%
            \def\PYGZob{\discretionary{}{\Wrappedafterbreak\char`\{}{\char`\{}}%
            \def\PYGZcb{\discretionary{\char`\}}{\Wrappedafterbreak}{\char`\}}}%
            \def\PYGZca{\discretionary{\char`\^}{\Wrappedafterbreak}{\char`\^}}%
            \def\PYGZam{\discretionary{\char`\&}{\Wrappedafterbreak}{\char`\&}}%
            \def\PYGZlt{\discretionary{}{\Wrappedafterbreak\char`\<}{\char`\<}}%
            \def\PYGZgt{\discretionary{\char`\>}{\Wrappedafterbreak}{\char`\>}}%
            \def\PYGZsh{\discretionary{}{\Wrappedafterbreak\char`\#}{\char`\#}}%
            \def\PYGZpc{\discretionary{}{\Wrappedafterbreak\char`\%}{\char`\%}}%
            \def\PYGZdl{\discretionary{}{\Wrappedafterbreak\char`\$}{\char`\$}}%
            \def\PYGZhy{\discretionary{\char`\-}{\Wrappedafterbreak}{\char`\-}}%
            \def\PYGZsq{\discretionary{}{\Wrappedafterbreak\textquotesingle}{\textquotesingle}}%
            \def\PYGZdq{\discretionary{}{\Wrappedafterbreak\char`\"}{\char`\"}}%
            \def\PYGZti{\discretionary{\char`\~}{\Wrappedafterbreak}{\char`\~}}%
        } 
        \newcommand*\Wrappedbreaksatpunct {%
            \lccode`\~`\.\lowercase{\def~}{\discretionary{\hbox{\char`\.}}{\Wrappedafterbreak}{\hbox{\char`\.}}}%
            \lccode`\~`\,\lowercase{\def~}{\discretionary{\hbox{\char`\,}}{\Wrappedafterbreak}{\hbox{\char`\,}}}%
            \lccode`\~`\;\lowercase{\def~}{\discretionary{\hbox{\char`\;}}{\Wrappedafterbreak}{\hbox{\char`\;}}}%
            \lccode`\~`\:\lowercase{\def~}{\discretionary{\hbox{\char`\:}}{\Wrappedafterbreak}{\hbox{\char`\:}}}%
            \lccode`\~`\?\lowercase{\def~}{\discretionary{\hbox{\char`\?}}{\Wrappedafterbreak}{\hbox{\char`\?}}}%
            \lccode`\~`\!\lowercase{\def~}{\discretionary{\hbox{\char`\!}}{\Wrappedafterbreak}{\hbox{\char`\!}}}%
            \lccode`\~`\/\lowercase{\def~}{\discretionary{\hbox{\char`\/}}{\Wrappedafterbreak}{\hbox{\char`\/}}}%
            \catcode`\.\active
            \catcode`\,\active 
            \catcode`\;\active
            \catcode`\:\active
            \catcode`\?\active
            \catcode`\!\active
            \catcode`\/\active 
            \lccode`\~`\~ 	
        }
    \makeatother

    \let\OriginalVerbatim=\Verbatim
    \makeatletter
    \renewcommand{\Verbatim}[1][1]{%
        \sbox\Wrappedcontinuationbox {\Wrappedcontinuationsymbol}%
        \sbox\Wrappedvisiblespacebox {\FV@SetupFont\Wrappedvisiblespace}%
        \def\FancyVerbFormatLine ##1{\hsize\linewidth
            \vtop{\raggedright\hyphenpenalty\z@\exhyphenpenalty\z@
                \doublehyphendemerits\z@\finalhyphendemerits\z@
                \strut ##1\strut}%
        }%
        \def\FV@Space {%
            \nobreak\hskip\z@ plus\fontdimen3\font minus\fontdimen4\font
            \discretionary{\copy\Wrappedvisiblespacebox}{\Wrappedafterbreak}
            {\kern\fontdimen2\font}%
        }%
        
        \Wrappedbreaksatspecials
        \OriginalVerbatim[#1,codes*=\Wrappedbreaksatpunct]%
    }
    \makeatother

    \definecolor{incolor}{HTML}{303F9F}
    \definecolor{outcolor}{HTML}{D84315}
    \definecolor{cellborder}{HTML}{CFCFCF}
    \definecolor{cellbackground}{HTML}{F7F7F7}
    
    \makeatletter
    \newcommand{\boxspacing}{\kern\kvtcb@left@rule\kern\kvtcb@boxsep}
    \makeatother

\usepackage{amssymb}

\usepackage{threeparttable}

\usepackage{tabularx}

\usepackage{color}

\usepackage{graphicx}

\usepackage{subfig}

\graphicspath{{images/}}

\usepackage{url}

\usepackage{pgfplots}

\usepackage{caption}

\usepackage{color}

\usepackage{amsmath}

\usepackage{lscape}

\usepackage{todonotes}

\usepackage{comment}

\usepackage{soul}

\usepackage[inline]{enumitem}

\newcommand{\eugenio}[1]{\textcolor{black}{#1}}

\newcommand{\dani}[1]{\textcolor{black}{#1}}

\newcommand{\paco}[1]{\textcolor{black}{#1}}

\newcommand{\new}[1]{\textcolor{black}{#1}}

\usepackage[breakable]{tcolorbox}
\usepackage{parskip} 

\usepackage{iftex}
\ifPDFTeX
	\usepackage[T1]{fontenc}
	\usepackage{mathpazo}
\else
	\usepackage{fontspec}
\fi

\newcommand{\nuria}[1]{\textcolor{black}{#1}}

\journal{Information Fusion}
\let\today\relax
\makeatletter
\def\ps@pprintTitle{%
    \let\@oddhead\@empty
    \let\@evenhead\@empty
    \def\@oddfoot{\footnotesize\itshape
         {} \hfill\today}%
    \let\@evenfoot\@oddfoot
    }
\makeatother

\begin{document}

\begin{frontmatter}


\title{\paco{Survey on Federated Learning Threats: concepts, taxonomy on attacks and defences, experimental study and challenges}}


\author[decsai]{Nuria Rodríguez-Barroso\corref{cor1}}
\ead{rbnuria@ugr.es}

\author[decsai]{Daniel Jiménez López}
\ead{dajilo@ugr.es}

\author[lsi]{M. Victoria Luz\'{o}n}
\ead{luzon@ugr.es}

\author[decsai]{Francisco Herrera}
\ead{herrera@decsai.ugr.es}

\author[decsai]{Eugenio Mart\'inez-C\'{a}mara}
\ead{emcamara@decsai.ugr.es}

\address[decsai]{Department of Computer Science and Artificial Intelligence, Andalusian Research Institute in Data Science and Computational Intelligence (DaSCI), University of Granada, Spain}

\address[lsi]{Department of Software Engineering, Andalusian Research Institute in Data Science and Computational Intelligence (DaSCI), University of Granada, Spain}

\cortext[cor1]{Corresponding author}



\begin{abstract}
\eugenio{Federated learning is a machine learning paradigm that emerges as a solution to the privacy-preservation demands in artificial intelligence. As machine learning, federated learning is threatened by adversarial attacks against the integrity of the learning model and the privacy of data \new{via a distributed approach to tackle local and global learning}. This weak point is exacerbated by the inaccessibility of data in federated learning, which makes harder the protection against adversarial attacks and evidences the need to furtherance the research on defence methods to make federated learning a real solution for safeguarding data privacy. In this paper, we present an extensive review of the threats of federated learning, as well as as their corresponding countermeasures, \new{attacks versus defences}. This survey provides a taxonomy of adversarial attacks and a taxonomy of defence methods that depict a general picture of this vulnerability of federated learning and how to overcome it. Likewise, we expound guidelines for selecting the most adequate defence method according to the category of the adversarial attack. Besides, we carry out an extensive experimental study from which we draw further conclusions about the behaviour of attacks and defences \new{and the guidelines for selecting the most adequate defence method according to the category of the adversarial attack. This study is finished leading to meditated learned lessons and challenges}}.

\end{abstract}


\end{frontmatter}

\section{Introduction} \label{sec:introduction}







Data-driven machine learning methods currently dominate artificial intelligence. This reliance on data allows us to stand out three artificial intelligence challenges. The former is the preservation of data privacy, since artificial intelligence methods process personal and sensitive data, such as health \cite{Al-Kuwari2021} and financial data \cite{Boissay2021}. Likewise, the growing interest in data privacy safeguarding is reflected in emerging legal frames such as the General Data Protection Regulation (GDPR) \cite{gdpr1}. The second challenge is related to the increasing availability of data, which, on the one hand, is furthering the progress of artificial intelligence \cite{GOMEZCARMONA2020670}, and, on the other hand, it arises new challenges related to its storage and processing that are even exacerbated when data stemmed from distributed sources, as in IoT scenarios \citep{zhang2021}. The latter challenge emerges from the need to distributively process data when it is not possible to transfer it to a central server, because of legal or regulatory restrictions, communication costs or other kind of technical limitations. Due to this distributed scenario, new difficulties appear linked to dissimilar data distributions from the same domain and the likely large size of data sources \cite{chenxin2017}.

\eugenio{Federated learning (FL) is a machine learning \paco{paradigm} proposed as a possible response to the three previous challenges, and especially for the demand of preserving data privacy, \paco{together with a distributed approach to tackle local and global learning} \cite{yang19}. FL aims at generating a collaboratively trained global learning model without sharing the data owned by the distributed data sources. Frequently, it requires a coordinator agent, which is in charge of managing the information exchange required to train the global learning model. In this way, the data is protected from unauthorised access, either by other data sources or the coordinator party.}


\eugenio{Machine learning is vulnerable to adversarial attacks mainly focused on impairing the learning model or violating data privacy \cite{huang2011}. Likewise, FL is exposed to the same jeopardy, since it is an specific machine learning setting. Some of those attacks are grounded in the maliciously manipulation of the training data \cite{dalvi2004}, which are inaccessible in FL and, then, we cannot rely on the use of data inspection techniques for detecting that altered data. Therefore, one of the weak points of FL is being exposed to adversarial attacks that may violate the integrity of the learning model or the privacy of data.}

\eugenio{The evidence that adversarial attacks are a weak point of FL is built upon the fact of the large volume of publications centred on the identification of vulnerabilities in the form of adversarial attacks \cite{bib:bagdasaryan18,DBLP:conf/uss/FangCJG20,Zhu2020deep,property:Zhang2020}, and on the corresponding large volume of defence proposals against to those attacks \cite{Sageflow,ozdayi2020defending,sun2020provable,DBLP:journals/corr/abs-2110-13864}. This effervescent quantity of publications is the cause of the publication of several survey works on adversarial attacks that attain to review and summarise the latest papers related to this weak point. These surveys lack of an holistic view of FL and the review of the defences against adversarial attacks, because of the following reasons: \begin{enumerate*}[label=(\arabic*)] \item most of them are only focused on one kind of adversarial attacks, namely there are surveys reviewing attacks to the federated model \cite{ji2019, DBLP:journals/corr/abs-1911-12560, Shejwalkar2021} or privacy attacks \cite{Enthoven2021, Asad2020, Mothukuri2021}, but any of them encompass both sort of attacks; \item the vast majority does not include any experimental \dani{study} \cite{Lyu2020_overview, 10.1007/978-3-642-40994-3_25, Lyu2020_threats, Jere2021, Bouacida2021}, so it is not possible to compare the strength of the attacks and the robustness of the defences in a common evaluation framework; and \item by default they only focus on horizontal FL ignoring vertical and federated transfer learning.\end{enumerate*}}

\paco{Due to the mentioned facts, we propose a new survey on FL threats, and additionally we provide several taxonomies on adversarial attacks and defences, an experimental study and a final discussion about lessons learned and challenges. This survey differs from previous ones due to the following contributions:}

\begin{enumerate}
    \item To provide a general picture of the field of adversarial attacks and defences by considering the threats to the learning model and to the integrity of the privacy of data.
    \item To review the threats and the defences of horizontal FL, vertical FL and federated transfer learning.
    \item To define a taxonomies of adversarial attacks and their corresponding defensive countermeasures. These two taxonomies encompass the different categories of adversarial attacks and defences, which will shed light in this crucial field of making FL a robust learning paradigm.
    \item To provide a guidelines for selecting the right defence category according to the threatening adversarial attack.
    \item To compare in a common evaluation framework the strength of the most relevant adversarial attacks, and the defence capacity of the most prominent defence methods.
    \item \paco{To expound some learning lessons stemmed from the literature review and the experimental study conducted.}
    \item \paco{To also expound their relations to the challenges in the field of adversarial attacks.}
\end{enumerate}


The rest of the paper is organised as follows: the following section introduces the \eugenio{propaedeutic} concepts \eugenio{necessary for this survey to be illustrative}. Section \ref{sec:model_attacks} presents the taxonomy of adversarial attacks in FL, while Section \ref{sec:defences1} expounds the taxonomy of defences against them. We conduct the experimental study in Section \ref{sec:experimental_study}. \paco{In Section \ref{guidelines} we provide the guidelines for selecting the right defence category. Finally, we discuss the lessons learned and challenges in Section \ref{sec:lessons} and \ref{sec:challenges}, and include some conclusions in Section \ref{sec:conclusions}}. 

\section{Background concepts on Federated Learning threats} \label{sec:background}



\eugenio{The concepts described throughout this paper require the knowledge of some propaedeutic concepts related to FL and its threats. Accordingly, w}e introduce FL and \eugenio{the categories of FL in} Section \ref{sec:fl}, we formally define differential privacy (DP) in Section \ref{sec:dp_definition}\eugenio{, since a considerable amount of defence methods are based on DP,} and we detail the categorization of the attacks in terms of the threat model in Section \ref{sec:threat_model}.

\subsection{Federated Learning}\label{sec:fl}


FL is a distributed machine learning paradigm with the aim of building a ML model without explicitly exchanging training data between parties \cite{yang19}. It consists in a network of clients or data owners $\{C_1, \dots, C_n\}$, who participate in two main processes:

\begin{enumerate}
\item \textit{Model training phase:} each client exchange information without revealing any of their data to collaboratively train a ML model, $\mathcal{M}_f$, which may reside at one client or may be shared between a few clients.

\item \textit{Inference phase:} clients collaboratively apply the jointly trained model, $\mathcal{M}_f$, to a new data instance.
\end{enumerate}

Both processes can be either synchronous or asynchronous, depending on the data availability of the clients and the trained model.

\dani{It must be highlighted the fact that privacy is not the only motivation of this paradigm, there should be a fair value-distribution mechanism to share the profit gained by the collaboratively trained model, $\mathcal{M}_f$.}

The distribution of characteristics of the data among clients in FL shapes the procedure to follow in the two main processes of FL, particularly we focus on the following distributions: \begin{enumerate*}[label={(\arabic*)}] \item clients share the feature space but not the sample space, \item clients share the sample space but not the feature space, and \item clients share only a small overlap in feature space. \end{enumerate*} These distributions allow us to present three categories of FL \cite{yang19} \dani{in terms of the feature space ($X$), the label space ($Y$) and the sample ID space ($I$) as follows}:


\paragraph{Horizontal Federated Learning (HFL)} \eugenio{In this scenario,} clients data share the feature and labels space, but differ in the sample space. Formally, we can define as: 

$$X_i = X_j, \; Y_i= Y_j, \; I_i \neq I_j, \; \forall D_i, D_j, \; i \neq j$$

where the feature and labels space of the clients $(i,j)$ is depicted by $(X_i,Y_i)$ and $(X_j,Y_j)$ and it is assumed to be the same, while the samples $I_i$ and $I_j$ are not the same. $D_i$ and $D_j$ depict the data of the clients $i$ and $j$.

\paragraph{Vertical Federated Learning (VFL)} In this scenario, clients share the sample space \dani{but neither the feature space nor the label space}. Formally, we can define as follows:

$$X_i \neq X_j, \; Y_i \neq Y_j, \; I_i = I_j, \; \forall D_i, D_j, \; i \neq j$$

\paragraph{Federated Transfer Learning (FTL)} This scenario is similar to the traditional transfer learning. The clients share neither the feature space, \dani{nor label space}, nor the sample space. Formally, we can define as follows:

$$X_i \neq X_j, \; Y_i \neq Y_j, \; I_i \neq I_j, \; \forall D_i, D_j, \; i \neq j$$

\begin{figure}[h!]
    \centering
    \includegraphics[]{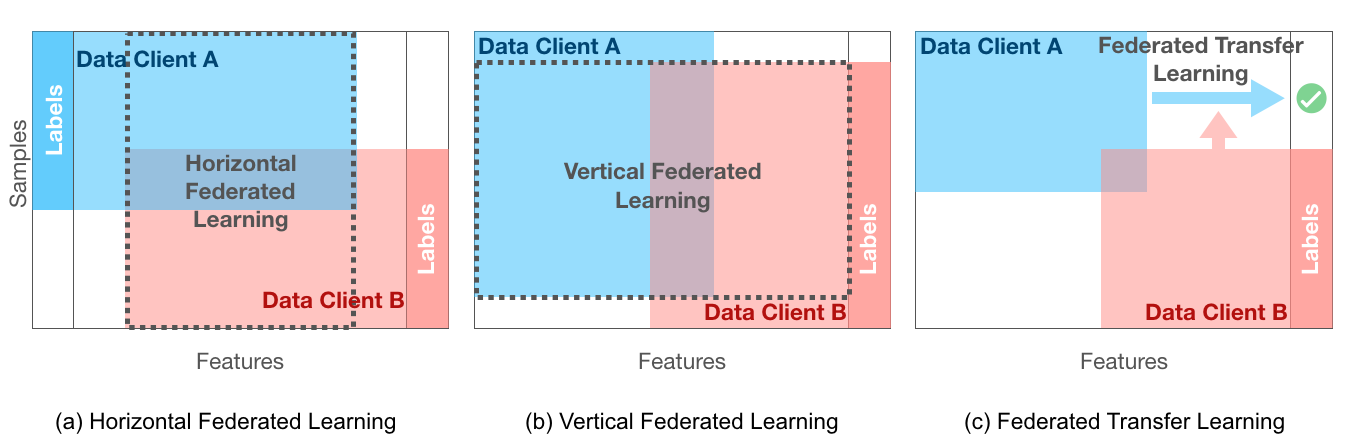}
    \caption{Representation of the different \dani{categories} in FL. \eugenio{Source \cite{yang19}.}}
    \label{fig:architectures}
\end{figure}

Although the feature space\dani{ and the label space} are not the same, in FTL there is a certain overlap or similarity, since the aim is to transfer knowledge from one client to another securely. FTL was presented in \cite{bib:yang19} and it represents higher difficulty than HFL and VFL, since it implies the use of techniques that preserve the data privacy. We represent the different \dani{categories} of FL in Figure \ref{fig:architectures}.

\eugenio{FL is a learning setting composed of a set of key elements. Since FL is a specific configuration of a machine learning environment, it} shares with \eugenio{machine learning} some \eugenio{of those} key elements, such as the data and the learning model. \eugenio{Nonetheless}, the particularities of FL make necessary additional key elements, such as clients and a learning coordinator that orchestrates the two main processes of FL. \eugenio{A detailed description of FL key elements focused on HFL is in \cite{rodrguezbarroso2020federated}, and here we describe the common ones to all the FL categories.}


\paragraph{\textbf{Data}}
It plays a central role in machine learning. In FL, data is distributed among the different clients according to two possibilities: (1) IID (Independent and Identically Distributed), when the data in each client is independent and identically distributed, as well as representative of the population data distribution; and (2) Non-IID (non Independent and Identically Distributed), when the data distribution in each client is not independent no identically distributed from the population data distribution. These data distributions are mainly relevant to HFL\dani{. In VFL and FTL categories, clients do not share \dani{neither the} feature space \dani{nor the} label space, and consequently the data distribution \eugenio{among clients} is relegated to a second place.} 

In most HFL scenarios, each client only stores the data generated on the client itself, ensuring the non-IID property of the global data. Moreover, even if the IID scenario were present, it would not be known because of the data privacy properties of FL. Hence, the non-IID scenario is the best choice and it represents a real challenge.






\paragraph{\textbf{Clients}}

Each client of a federated scenario plays a key role in a federated paradigm, as a data owner and as a part of the distributed scheme. Typical clients in FL could be servers, smartphones, IoT devices, connected vehicles, hospitals, banks or insurance companies. Privacy is not their only motivation, they also want to profit from the \textit{model training phase}. As a consequence, a reward mechanism is expected,  such as owning the collaboratively trained model, $\mathcal{M}_f$, in HFL or the outputs of the \textit{inference phase} in VFL and FTL.


\paragraph{\textbf{Learning coordinator}}

\dani{The learning coordinator} orchestrates the communication among the clients in the two main processes of FL. While it is not strictly necessary, when present, it also plays the role of a trusted authority. In VFL, \eugenio{the learning coordinator receives and combines partial updates from clients and shares the corresponding part of the combined update with each client} in the \textit{model training phase}. Moreover, in the \textit{inference phase} it helps to perform the inference by combining the outputs of each client as the collaboratively trained model, $\mathcal{M}_f$, is split among \dani{them}. In contrast to VFL, in HFL the learning coordinator is usually known as th federated server and it only participates in the \textit{model training phase}: \begin{enumerate*}[label={(\arabic*)}] \item receiving the trained parameters of the local models, \item aggregating the trained parameters of each client model using federated aggregation operators and \item updating every learning model with the aggregated parameters. \end{enumerate*}. \dani{Moreover,} the \textit{inference phase} is not performed in a collaborative way as the collaboratively trained model, $\mathcal{M}_f$ is stored in each client and in the federated server.

\subsection{Differential Privacy}\label{sec:dp_definition}

DP allows retrieving information, rigorously bounding the harm caused to individuals whose sensitive data are stored in the database \cite{TCS-042, Dwork2006}. Basically, it hides the presence of an individual in the database. To achieve this, DP adds random noise to the outputs. Such noise is calibrated to the magnitude of the largest contribution that can be made to the output by an individual. It is important to note that DP assumes that the adversary owns arbitrary external knowledge.

DP is the key property used to provide a certain level of privacy to any sensitive data access, in a way it is both, secure and measurable. It is secure because it has a theoretical background which supports it. It is measurable as every access to private data has a privacy cost either in terms of $\epsilon$ or in terms of $(\epsilon, \delta)$.

This interpretation naturally leads to define the \textit{distance between databases}: two databases $x,\,y$  are said to be $n$-neighbouring if they differ by $n$ entries. In particular, if the databases only differ in a single data element ($n=1$), the databases are simply addressed as \textit{neighbouring}. 

\paragraph{\textbf{Differential Privacy definition}} A database access mechanism\eugenio{,} $\mathcal{M}$, preserves $\epsilon$-DP if for all neighbouring databases $x,\,y$ and each possible output of $\mathcal{M}$, represented by $\mathcal{S}$, it holds that:
\begin{equation}\label{eq_dp_definition}
	P[\mathcal{M}(x) \in \mathcal{S}] \leq e^{\epsilon} \   P [\mathcal{M}(y) \in \mathcal{S}] 
\end{equation}
If, on the other hand, for $0<\delta< 1$ it holds that:
\begin{equation}\label{eq_dp_weaker_definition}
P[\mathcal{M}(x) \in \mathcal{S}] \leq e^{\epsilon} \ P [\mathcal{M}(y) \in \mathcal{S}] + \delta
\end{equation}
then the mechanism possesses the property of $(\epsilon, \delta)$-DP, also known as approximate DP. 

In other words, DP specifies a ''privacy budget'' given by $\epsilon$ and $\delta$. The way in which it is spent is given by the concept of privacy loss. The privacy loss allows us to reinterpret both, $\epsilon$ and $\delta$ in a more intuitive way:

\begin{itemize}[noitemsep]
    \item $\epsilon$ limits the quantity of privacy loss permitted, that is, our privacy budget. 
    \item $\delta$ is the probability of exceeding the privacy budget given by $\epsilon$, so that we can ensure that with probability $1-\delta$, the privacy loss will not be greater than $\epsilon$.
\end{itemize}

DP has some interesting properties, which makes it even more appealing in a privacy context.

\begin{enumerate}
    \item \textbf{DP is immune to post-processing}. if an algorithm protects an individual's privacy, then there is not any way in which privacy loss can be increased.
    
    \item \textbf{DP can be used to protect the privacy of groups}. Let $\mathcal{M}$ be a $\epsilon$-differentially private mechanism, then $\mathcal{M}$ is  $K \epsilon$-differentially private for groups of size $K$.
    
    \item \textbf{DP mechanisms can be composed multiple times and remain differentially private}. Let $\mathcal{M}_1$ and $\mathcal{M}_2$ be $\epsilon_1$-differentially private mechanism and $\epsilon_2$-differentially private mechanism, respectively. Then, their composition output given by the concatenation of the output of  $\mathcal{M}_1$ and  $\mathcal{M}_2$ over the same input is $\epsilon_1+\epsilon_2$-differentially private

\end{enumerate}

\subsection{Threat Model}\label{sec:threat_model}




\dani{Threat models in machine learning are structured representation of information\eugenio{,} which help to identify and define potential security issues. They can be defined in terms of the information available and the \dani{scope} of action of the attacker. In this regard, we define the following set of mutually exclusive terms that allow us to \dani{define} the FL threat model}. 

\paragraph{Insider vs. Outsider} One of the key elements of any distributed system is the communication between different parts. \eugenio{The communication is} very vulnerable\eugenio{,} since it can be compromised by agents from outside the learning system, which are known as outsider attackers. When the attack is carried out by one of the participants in the distributed system, either one or more clients, or the server, it is known as an insider attacker. Clearly, the scope of the two attacks is very different: insider attacks \eugenio{are more harmful} and may be aimed at modifying the behaviour of the model or inferring valuable information from other clients, while those carried out by outsiders are usually aimed only at inferring information about the data or the resulting learning model. Outsider attacks mainly focus on sniffing information of the communication channels between the involved agents. They are either side-channel attacks, when the attacker gains information from the implementation of the FL scenario, or man-in-the-middle attacks, when the attacker intercepts the communication channel by disguising herself as the receiver part. Both attacks are related to the protocols used to establish communication and their implementation.

\eugenio{We focus on insider attacks, in which we highlight the following categorisations:}
\begin{itemize}
    \item \textbf{Byzantine attacks}. \eugenio{They consist} in sending arbitrary updates to the server, so it compromises the performance of the global learning model.
    \item \textbf{Sybil attacks}. They consist of collaborative attacks, either by several attackers joining together or by simulating fictitious clients in order to be more disruptive.
\end{itemize}

\paragraph{Client vs. Server} Regarding insider attacks, \dani{in HFL} it is natural to differentiate between two types of attacks, depending on whether they are carried out by a client or by a server. The main point of difference lies in the amount of information available. While the attacks carried out by clients only have information of one or several clients, the server holds information about the model architecture and the updates of the clients in each round of learning. Even, in cryptographic implementations of the federated communication among the federated server and the clients, the server owns more information than the clients\eugenio{,} as it is the only one with enough knowledge to \eugenio{decipher} the models.

\paragraph{Attacker knowledge} In centralised settings, the white-box attacker has full access to the target model, including the model architecture, the parameters and its internal state. In contrast, the black-box attacker does not have any access to the target model and additionally, she might have some additional information about the architecture of the target model or its training procedure. These two classifications of attacker knowledge are too general to represent every type of attacker\eugenio{,} because there is no middle ground to consider attackers whose knowledge in the black box setup is too restricted\eugenio{,} and in the white box setup is not enough constrained. To address this issue, a grey-box attacker was introduced in \citep{Truex2018}, which is a black-box attacker with some specific statistical knowledge not publicly available that concerns her victim. This description of attacker knowledge is tailored for a centralised learning setting, and as a consequence it does not fit other learning settings as the attack surface changes. In a FL system, white-box, grey-box or black-box attackers can be any \eugenio{node, either the clients or the server}. Moreover, the exposed attack surface is greater than in centralised settings. Most attacks are related to the data owned by the clients and the communication among the federated server and the clients, therefore, we also require including \dani{the information available regarding the federated training process} and to the client's private data. In order to address such requirements, we define the following classification of the attacker's knowledge suited for HFL and VFL:

In a standard HFL system, an attacker which owns a client has \textit{client-side knowledge}:

\begin{itemize}[noitemsep]
    \item White-box access to the aggregated model.
    \item White-box access to the client's locally trained model.
    \item Access to the owned client's dataset.
\end{itemize}

If the attacker has access to local data of other clients or their labels, she has \textit{extra client-side knowledge}.

An attacker which owns a federated server has \textit{server-side knowledge}:
\begin{itemize}[noitemsep]
    \item White-box access to the aggregated model after each communication round.
    \item White-box access to trained models shared by the clients or, alternatively, access to their gradients.
    \item The identifiers of the clients aggregated in each communication round.
    \item The labels owned by each client and, optionally, the size of their dataset.
\end{itemize}

In a standard VFL system, an attacker which owns a client has \textit{party-side knowledge}:

\begin{itemize}[noitemsep]
    \item White-box access to the parameters related to the features of the owned client.
    \item Access to the client's private dataset.
    \item The partial output of the parameters, when an inference is requested.
\end{itemize}

Additionally, if the attacker has access to information related to the features of the other clients, she has \textbf{extra party-side knowledge}.

An attacker which owns the learning coordinator in a VFL system has \textit{third party-side knowledge}:

\begin{itemize}[noitemsep]
    \item The gradients shared by each client.
    \item The computed loss.
    \item The partial output of each client, when an inference is requested.
\end{itemize}

\dani{If only a subset of the specified knowledge is available to the attacker, then she has \textit{partial} knowledge, and we specify the content of that subset of knowledge. Moreover, defences are expected to reduce the attacker knowledge, therefore in the presence of a defence an attacker is expected to have \textit{partial} knowledge.}

In both HFL and VFL systems, if the attacker only have access to the outputs of the federated model, she has \textit{outsider-side knowledge}.

We highlight the fact that the categories stated are not mutually exclusive, that is, an attacker can own multiple types of knowledge at the same time. Realistic attack scenarios tend to require lesser attacker knowledge, while more complex and specific attacks require knowledge from multiple participants of a FL task.

\paragraph{\textbf{Honest-but-curious vs. Malicious}} A malicious (or active) attacker tries to interfere in the training process of the learning model with the aim of corrupting the target model, for example, damaging its performance or injecting a secondary task. On the contrary, an honest-but-curious (or passive) attacker does not interfere in the training process and follows the federated learning protocols, but try to obtain private information about other clients from the received information.

\paragraph{\textbf{Collusion vs. No-collusion}} The collusion threat lies in the fact that the attacker who controls more clients has more power in a distributed system. There are two collusion types: \begin{enumerate*}[label=(\arabic*)] \item server-participants, in which the attacker controls some benign participants and the server\eugenio{, and it} aims to infer information about the rest of the clients; and \item participant-participant, in which the attacker controls a fraction of the benign clients and aims to infer information about benign clients, the server or to harm the learning model. \end{enumerate*}

\section{Adversarial Attacks in Federated Learning\paco{: Taxonomies}} \label{sec:model_attacks}

\dani{Adversarial attacks represent one of the more challenging problems in FL, due to the large number of existing attacks and the difficulty of defending against them. Moreover, the distribute nature of FL makes it vulnerable to wide variety of adversarial attacks aiming at different objectives and using different ways to achieve these objectives. Due to this wide variety in the nature and target of attacks, it is difficult to establish a common taxonomy for all types of adversarial attacks. For this reason, we propose the first broadly classification by differentiating between:}

\begin{itemize}[noitemsep]
    \item \dani{\textbf{Attacks to the federated model}, which aim at modifying its behaviour.}
    \item \dani{\textbf{Privacy attacks}, whose purpose is to infer sensitive information from the learning process.}
\end{itemize}

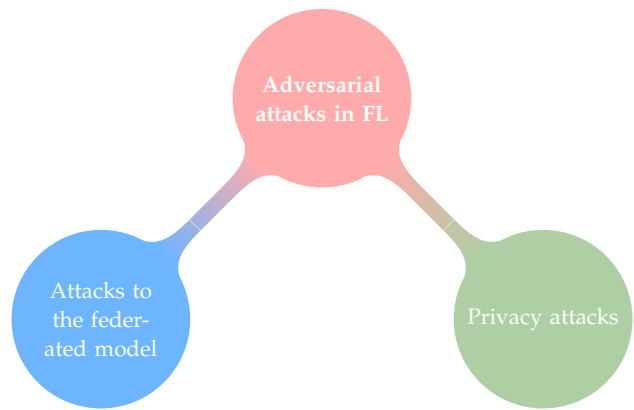
\begin{figure}[h!]
\centering
\input{main_categorisation.tex}
\caption{First, categorisation of the adversarial attacks in FL into two broad categories: attacks to the federated model and privacy attacks.}
\label{fig:main_categorisation}
\end{figure}

\dani{In Figure \ref{fig:main_categorisation} we represent this first categorisation of the adversarial attacks in FL.}

\dani{Once this initial classification into these two main categories of attacks has been established, we further examine each category by proposing a taxonomy based on different criteria and review the most relevant works on each topic. In Section \ref{sec:attacks_to_model} we focus on attacks to the federated model and the Section \ref{sec:inference_attacks} is dedicated to the privacy attacks.}

\subsection{Adversarial attacks to the federated model}\label{sec:attacks_to_model}

\dani{One of the main limitation of FL, and more specifically of the HFL, in terms of adversarial attacks, is that clients have the ability to harm the model by sending poisoned updates, while the server cannot inspect the training data stored on the clients. This fact makes the adversarial attacks to the federated model become one of the most significant challenges in FL.}

\dani{In general, these attacks are carried out by clients and the white-box feature of these attacks correspond to the situation in which the attacker has client-side knowledge, either there are one or several adversarial clients (attackers). In some situations attackers are considered to have access to more white-box information, for example about the aggregation mechanism used on the server, which is not a realistic situation. We therefore highlight those attacks that only require information from the adversarial client.}

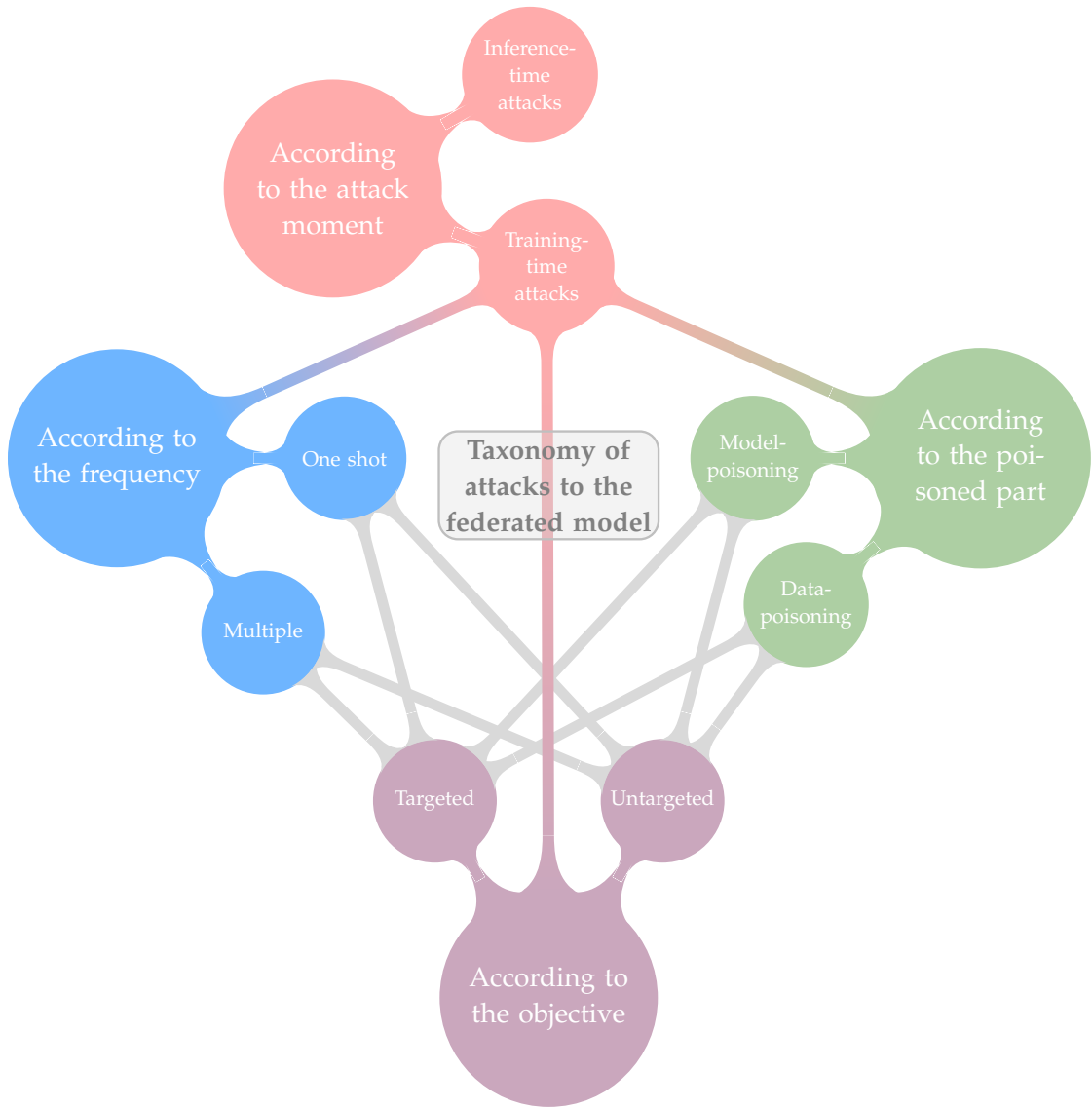
\begin{figure}[!t]
\centering
\input{attacks_to_the_model.tex}
\caption{\eugenio{Representation of the attack taxonomies to the federated model according to the different criteria. \dani{The grey links represent the possibility of combination of both categories. For the sake of clarity, we don't show redundant connections between categories already connected with other links.}}}
\label{fig:attacks_to_model}
\end{figure}

Within this broad category, we propose \eugenio{a taxonomy that encompasses a range of attacks according to different criteria, which we depict in Figure \ref{fig:attacks_to_model}}. Thus, each type of attack in the literature belongs to \dani{four} different categories, one for each criterion. \eugenio{From the main taxonomy, we additionally propose four more taxonomies linked to each criterion, namely:} \dani{(1) the attack moment in Section \ref{sec:attack_moment}, (2) the objective in Section \ref{sec:objective}, (3) the poisoned part of the FL scheme in Section \ref{sec:poisoned} and (4) the frequency in Section \ref{sec:frequency}.}



\subsubsection{Taxonomy according to the attack moment}\label{sec:attack_moment}

We present the taxonomy according to the time at which the attack is carried out, which completely determines the ability of the attack to influence the federated model. \dani{W}e classify the following two types of attacks:

\paragraph{\textbf{Training time attacks}} The training time phase includes from data collection and data preparation to model training. These attacks are carried out during this phase, either continuously or as a single attack. They are the most common in the literature since they have the ability to modify the federated model that is still being trained \cite{bib:bagdasaryan18, fung2018, bhagoji2018model} and to infer some information from training data \cite{Nasr_2019} (see Section \ref{sec:inference_attacks}). 

\paragraph{\textbf{Inference time attacks}} These attacks are carried out in the \dani{\textit{inference phase}} \eugenio{ when the model has been trained.} They are called evasion or exploratory attacks \cite{10.1007/978-3-642-40994-3_25}. Generally, the objective is not to modify the trained model, but to produce wrong predictions or to collect information about the characteristics of the model.
    
\subsubsection{Taxonomy according to the objective}\label{sec:objective}

\dani{The most widely used categorisation in the literature, which makes it the most significant criteria is based on the target of the attack}. Although all the attacks in this section are gathered under the scope of modifying the model, the modifications can be quite diverse. We distinguish two broad groups depending on the target of the attack:

\paragraph{\textbf{Targeted or backdoor attacks \cite{Chen2017TargetedBA, DBLP:journals/corr/abs-1911-07963, bib:bagdasaryan18}}}  The main task is to inject a secondary or backdoor task into the model. In other words, a backdoor attack is successful as long as it succeeds in preserving its performance in the original task while injecting a second task. These attacks are very stealthy, since they generally do not affect the performance of the original task \cite{DBLP:journals/corr/abs-1811-12470}, which makes them hard to detect. \dani{Note that although they do not pose a danger to the FL main task, they do represent a danger to the integrity of the system, since the attacker takes advantage of the federated infrastructure to perform a certain backdoor action, representing a security breach.} 
The nature of such attacks is \dani{broad, given the great variety of secondary tasks}. \dani{We present a taxonomy based on different criteria, which is shown in Figure \ref{fig:backdoor_attacks}, with the following categories being the most} \dani{frequent}:
\begin{itemize}
    \item \textit{Input-instance-key strategies}. The objective is that the model labels specific input examples with a specific target label different from the original one. \dani{For example, in a face recognition system that allows access to a house, to identify five specific people from the input set, who originally did not have access (negative label as origin label) as people who can access (positive label as a target label)}. \dani{Some works which implement this kind of the attack are \cite{ji2019} where the authors analyse the impact of different attacks scenarios, \cite{wang2020} where the authors prove that you can really backdoor FL even using existing defences and \cite{DBLP:journals/corr/abs-2004-10020} where the aim is to present the data-poisoning attacks}.
    \item \textit{Pattern-key strategies}. The objective is that the model associates a particular pattern in an input sample with a particular target label. \dani{For example, in the face recognition system above, to allow access to any person wearing a polka-dot bow. In this way the system would identify the pattern "polka-dot bow" with the target label (positive label)}. In practice, a simple pattern of a cross or similar mark are chosen for association with a target label \cite{DBLP:journals/corr/abs-1911-07963, bib:bagdasaryan18}. In Figure \ref{fig:pattern-key}, we depict an attack using the pattern-key strategy of associating the blue cross with the target label. 
    
    \begin{figure}[!t]
        \centering
        \includegraphics{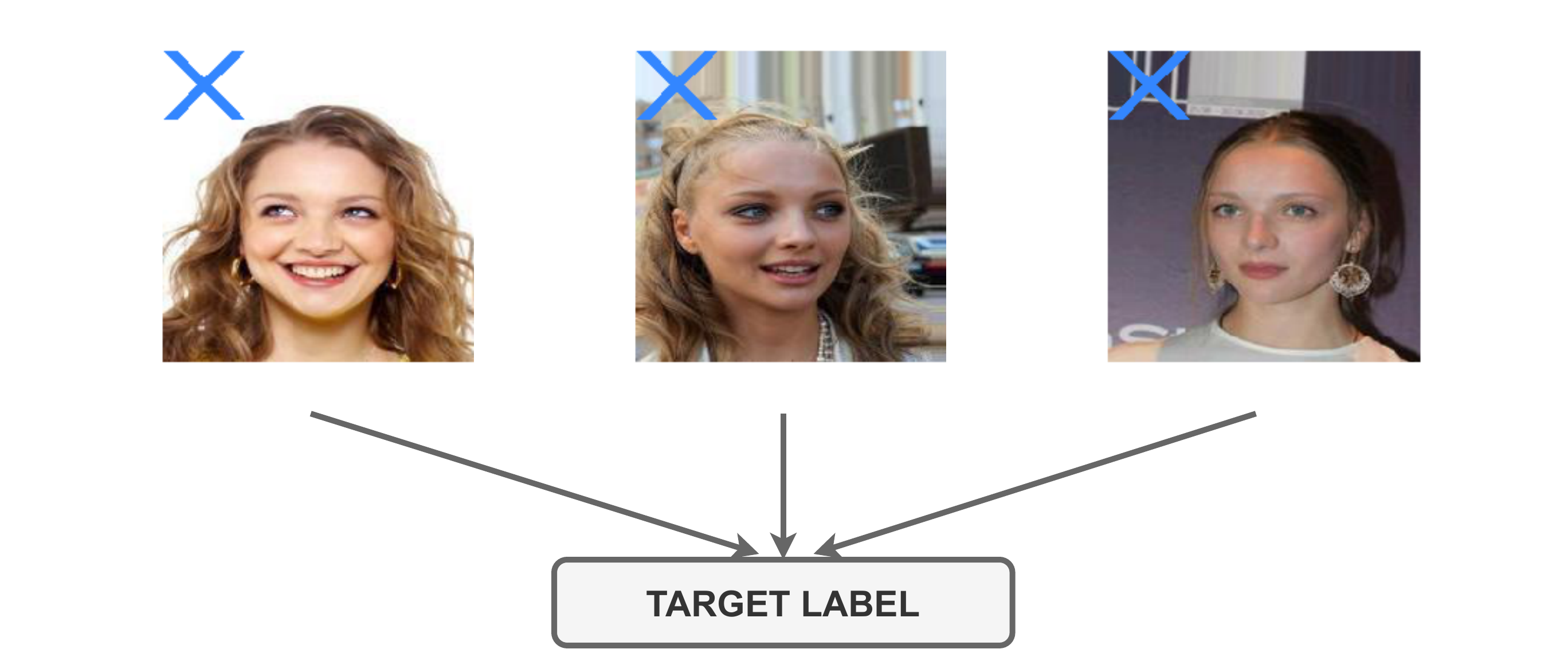}
        \caption{Representation of an attack using pattern-key strategy based on associate the blue cross with some prefixed target label.}
        \label{fig:pattern-key}
    \end{figure}
    
    Additionally, these attacks can also be categorised according to different criteria about the injected pattern \dani{as shown in Figure \ref{fig:backdoor_attacks}}. 

    \begin{figure}[!t]
    \centering
    \input{backdoor_attacks}
    \caption{Representation of the taxonomy of backdoor attacks.}
    \label{fig:backdoor_attacks}
    \end{figure}
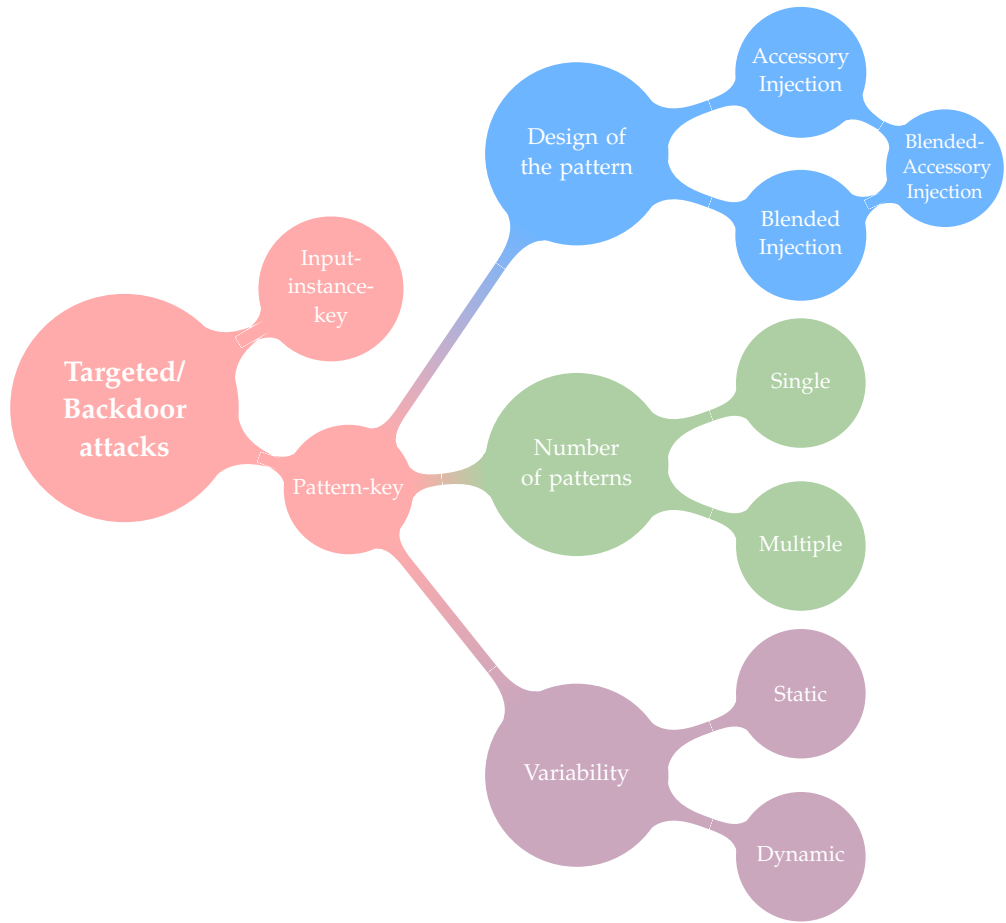
    
    Regarding the design of the pattern \dani{in \cite{Chen2017TargetedBA} the authors introduce the following terminology with the aim of classifying pattern attacks. Although this classification is not usually specified in other FL work, it is common in ML, and we believe it would be useful to use this notation in FL attacks as well.}:
    \begin{itemize}
        \item \textit{Blended injection strategy.} This strategy generates backdoor instances by blending a benign input instance with the key pattern using a blend ratio. The pattern can be any image, for example cartoon images or randomly generated patterns. The main limitation is that this mechanism requires to modify the entire sample during both training and testing, which may not be feasible.
        \item \textit{Accessory injection strategy.} This attack arises as a solution to the main limitation of the Blended injection strategy and proposes to generate backdoor images adding patterns to some regions of the original images. They are equivalent to wearing an accessory \dani{in real life}.
        \item \textit{Blended accessory injection strategy.} It takes advantage of both strategies by combining the accessory and the blended approach. 
    \end{itemize}
    Regarding the number of patterns:
    \begin{itemize}
        \item \textit{Single pattern attack}. It refers to when all adversarial clients inject the same pattern into the model. They are usually more successful as they are a collective attack on the same target, but at the same time easier to identify on the server. \dani{This situation is the most common one and some works such as \cite{bib:bagdasaryan18, Chen2017TargetedBA} where the authors focus on presenting the vulnerabilities of FL to such attacks\eugenio{,} or  \cite{ozdayi2020defending} where the aim is to propose a defence mechanism against them implement single pattern attacks.}
        
        \item \textit{Multi-backdoor attack} \cite{bib:bagdasaryan18}. It is composed of several coordinated adversarial clients (sybils)\eugenio{,} where each of them injects a different pattern or part of a common pattern to the model \cite{Xie2020DBA}. On the contrary, they are more difficult to detect on the server because the distribution of the pattern across clients enhances \dani{the stealth. Though,} it is more complicated for clients to inject backdoor tasks into the model, due to the diversity of secondary tasks.
    \end{itemize}
    
    Regarding the variability over time of the pattern:
    
    \begin{itemize}
        \item \textit{Static attack}. When the pattern of the attack is maintained over time regardless of the frequency of the attack. This situation is the most common one and some works \dani{cited before} such as \cite{bib:bagdasaryan18, Chen2017TargetedBA, ozdayi2020defending} implement static attacks.
        \item \textit{Dynamic attack}. The pattern changes over time, which is a challenge both for the defences, as the pattern to be identified changes, and for the adversarial clients, as they have to continuously adapt to new secondary tasks increasing the computation required. \citet{salem2021dynamic} propose to use meta-learning in order to speed up the adaptation of clients to the new backdoor tasks, and design a "symbiosis network" in which the clients weight the update of the model weights with the global model, instead of \eugenio{completing} replacement in order to maintain the performance on the backdoor tasks.
    \end{itemize}
\end{itemize}

Some works question the strength of backdoor attacks, since the most naive approaches are mitigated by simple defences \cite{DBLP:journals/corr/abs-1911-07963}. However, the potential of these attacks is shown in \citet{wang2020}, where they demonstrate that poisoning samples belonging to the tails of the data distribution is \dani{enough} to compromise the \dani{federated} global model. In addition, \citet{DBLP:journals/corr/abs-2007-03608} show that even attackers with no access to training labels can inject backdoor attacks in feature-partitioned collaborative learning. In conclusion, preliminary studies show that backdoor attacks are a real threat to FL, which further increases the interest in this \dani{research area}.

\paragraph{\textbf{Untargeted attacks} \cite{fang2019local, NIPS2017_6617}} As opposed to targeted attacks, the only goal of untargeted attacks is to impair the performance of the model on the original task. The most extreme scenario is known as \textit{Byzantine attacks} \cite{10.1145/357172.357176, hu2021challenges}, in which adversarial clients share randomly generated model updates or train over randomly modified data, generating random model updates as well. Clearly, these attacks are inherently less stealthy than targeted attacks, and can be detected merely by analysing the performance of the local models updates on the server, although it is sometimes difficult to differentiate them from clients with very particular training data distributions.

It is worth mentioning the \textbf{\textit{free-riders attacks}}. It is common in FL systems for clients to be awarded rewards for participation, as they provide crucial and necessary information. These rewards may tempt some clients to pretend that they are participating in the local training process and send updates to their models. To this end, they generate their "model updates" randomly resulting in the same effect as Byzantine attacks \cite{DBLP:journals/corr/abs-2006-11901}.

\subsubsection{Taxonomy according to the poisoned part of the FL scheme}\label{sec:poisoned}

Most training-time model attacks are based on poisoning client's information in order to corrupt the global learning model. Depending on which part of the client's information is poisoned, we \dani{differentiate between data-poisoning and model-poisoning attacks, and we refer both attacks as poisoning attacks. Figure \ref{fig:poisoned} shows the taxonomy presented in the rest of the section. In the following, we detail each one of them}:

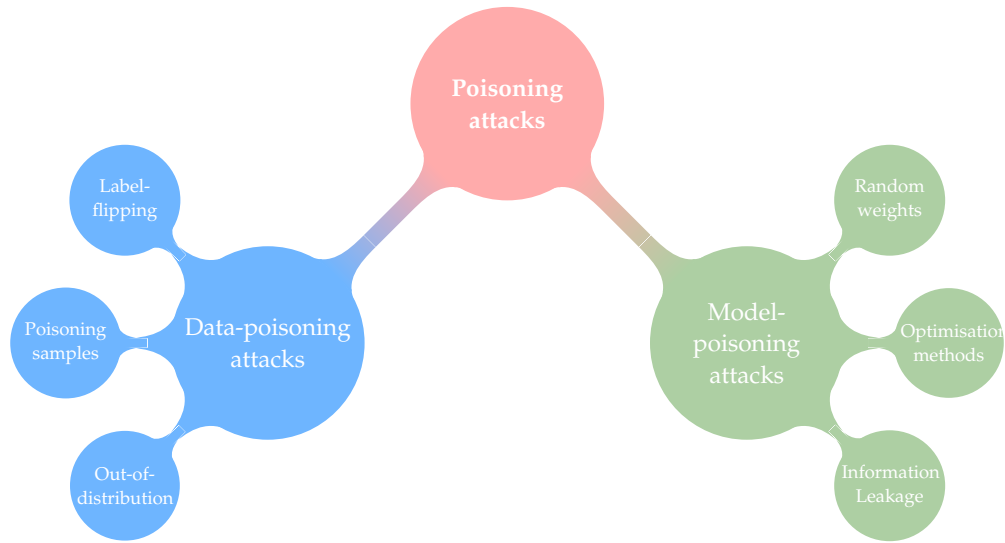
\begin{figure}[h!]
    \centering
    \input{poisoned}
    \caption{Representation of the taxonomy of backdoor attacks according to the poisoned part of the FL scheme.}
    \label{fig:poisoned}
\end{figure}

\paragraph{\textbf{Data-poisoning attacks} \cite{DBLP:journals/corr/abs-2007-08432, 8975792}} The attacker is assumed to have access to the training data of one or more clients and to be able to modify it. Depending on the characteristics of the poisoning, we \eugenio{distinguish} between the following attacks:

\begin{itemize}
    \item \textit{Label-flipping attack} \cite{9650669}. This attack consists of modifying the labels of a portion of the training data. It can be either targeted, by exchanging some specific labels \cite{DBLP:journals/corr/abs-2007-08432}, or untargeted \cite{hu2021challenges}, by random label shuffling.
    \item \textit{Poisoning samples attack}. Unlike the previous one, this attack consists of modifying part of the training data samples. The poisoning can be of different types, such as including patterns in the samples and associate it with some target class, or normalizing the samples and adding uniform noise with the aim of impairing the performance of the model. In recent years, the use of Generative Adversarial Nets (GANs) \cite{NIPS2014_5ca3e9b1} to generate these poisoned samples has become popular, to maximize the target of the attack on the one hand\dani{, and on the other hand,} to maximize the disguise of the attack to overcome the possible defences of the server on the other \cite{9194010}. A further clear example is the case of the attack proposed in \cite{8887357}, which consists in: (1) the attacker first behaves as a benign client and trains a GAN to mimic prototypical samples of other benign clients and, then, (2) the attacker generates the poisoned samples using these generated samples in order to compromise the global model by sending scaled poisoning updates as their local model updates.
    \item \textit{Out-of-distribution attack}. This attack is similar to the poisoning samples attacks\eugenio{, although they differ} in that the poisoned training samples are not modifications of the original ones, but samples from outside the input distribution \cite{fort2021exploring}. It is possible to use \dani{either samples from another domain with the same characteristics or samples made of random noise.} 
\end{itemize}

One of the key factors for the success of a data poisoning attack is the proportion of adversarial clients, and the amount of data they poison. In \cite{8975792}, they experiment with different data-poisoning attacks and conclude that: (1) the attack success increases linearly with the number of poisoned samples; (2) the increment of the number of attackers could improve the attack success without changing the total number of poisoned samples; and (3) the attack success increases faster with the number of poisoned samples \dani{than} when there are more attackers involved. 

The goal of most data-poisoning attacks is to impair the global model and thus the \eugenio{local} models of all clients. However, it is also possible that the goal of the attackers is not to impair of the \eugenio{local} models, but only a specific subset of them. In \citet{DBLP:journals/corr/abs-2004-10020}, they define a set of target nodes as those nodes (clients or server) to be compromised by the attack. According to this definition, we may differentiate between the following three types of data-poisoning attacks depending on \dani{the access level} the attackers have to the target nodes:
\begin{itemize}
    \item \textit{Direct attack}. The attackers have access to target nodes, so they inject poisoning samples directly on them.
    \item \textit{Indirect attack}. The attackers have no access to target nodes, so they have to employ further mechanisms such training themselves (in case they are clients) on the poisoned samples to poison the global model, which will then shared with the target clients.
    \item \textit{Hybrid attack}. When the attackers combine both previous attacks.
\end{itemize}
In the vast majority of the attacks in the literature, the attackers are supposed to have access to the target nodes, so the most common attacks are direct attacks.

\paragraph{\textbf{Model-poisoning attacks}} These attacks consist of directly poisoning the model updates sent by the clients to the server. Although data-poisoning attacks naturally lead to model-poisoning attacks, in this section we focus only on those attacks that directly modify the local update weights. Depending on how these model weights are generated, we distinguish between:

\begin{itemize}
    \item \textit{Random weights generation}. These attacks are based on generating the model weights as a vector of randomly generated values of the same dimension as the model weights received from the server. \eugenio{Two specific examples are: \begin{enumerate*}[label=(\arabic*)] \item the \textit{random weights attack} \cite{DBLP:journals/corr/abs-1911-12560}, in which an interval [-R,R] is inferred from the global learning model and the weights randomly generated in that interval; and \item the \textit{Gaussian attack} \cite{DBLP:conf/uss/FangCJG20}, a white-box attack, which chooses as model weights a sample from the Gaussian distribution resulting of the other clients' model updates.\end{enumerate*}
    \dani{By construction, the random weights attacks are more harmful while being easier to detect, so depending on the scenario it would be more dangerous one or the other.}
    }
    \item \textit{Optimization methods}. They consist of maximizing performance in the backdoor task, while minimizing the differences of the poisoned model with respect to the shared model by the server in the last round, thus maximizing effectiveness and stealth. This challenge is approached as a multi-objective optimization problem \cite{koh2018stronger}. This methodology is quite versatile and can be used to attack in special situations. For example, it is widely used to attack specific defences by introducing new criteria to be optimized \cite{DBLP:conf/uss/FangCJG20} in order to overcome defences discarding conditions \dani{specific to each defence}. In addition, in \cite{koh2018stronger} they also prove that regularization techniques decrease the impact of \dani{the training data} in the resulting model. For that reason, they propose to train adversarial clients without any regularization mechanism in order to increase the impact of the poisoned samples. \dani{This kind of attack is probably the most efficient approach to perform targeted attacks on the model.}
    \item \textit{Information leakage}. A particular use case of model-poisoning attacks in FL \eugenio{is} information leakage, where the objective is not to compromise the global model, but the communication among the attackers through a secure protocol \cite{10.1145/3411501.3419423}. It consists in the fact that certain clients are coordinated in such a way that they know common rules and by modifying small parts of the model weights they can communicate. In \cite{10.1145/3411501.3419423} is proposed to adjust the training data strategically so that the weight of a particular dimension in the global model will show a pattern known by the rest of the malicious clients. Along very similar lines, \citet{DBLP:journals/corr/abs-2104-10561} put forward a novel attacker model aiming at turning FL systems into covert channels to implement a stealth communication infrastructure by means of modifying certain bits of the models.

\end{itemize}

In FL, with the assumption that the proportion of adversarial clients is significantly lower than that of benign ones, the effect of the attack is expected to be dissipated in the aggregation. Therefore, \textit{model-replacement} techniques \cite{DBLP:journals/corr/abs-1811-12470, DBLP:journals/corr/abs-1911-07963, bib:bagdasaryan18} are used, which consist of weighting the contribution of adversarial clients using boosting techniques in order to replace the aggregated model with its local updates. Formally, if we consider the update of the global model in the learning round $t$ is computed as follows in Equation \ref{agg}:

\begin{equation}\label{agg}
    G^{t} = G^{t-1} + \frac{\eta}{n} \sum_{i=1}^n (L_i^{t} - G^{t-1}),
\end{equation}

where $G^t$ is the aggregated model at the learning round $t$, $L_i^{t}$ the model update of the client $i$ at the learning round $t$, $n$ the number of clients participating in the aggregation and $\eta$ the server's learning round. 

In this context, we consider the local model update of the adversarial client trained on the poisoned training data as follows in Equation \ref{model-replacement}:

\begin{equation}\label{model-replacement}
    \hat{L}_{adv}^{t} = \beta (L_{adv}^{t} - G^{t-1}),
\end{equation}

where $\beta = \frac{n}{\eta}$ is the boost factor. After that, replacing Equation \ref{model-replacement} in Equation \ref{agg} we have\footnote{We assume that the adversarial client is client 1.}

\begin{equation}\label{replacement}
    G^{t} = G^{t-1} + \frac{\eta}{n} \frac{n}{\eta}(L_{adv}^{t} - G^{t-1}) + \frac{\eta}{n}\sum_{i=2}^n (L_i^{t} - G^{t-1}).
\end{equation}

According to the definition of FL \cite{bib:konecny16federatedlearning}, eventually the FL model will converge to a solution, so we can assume that $L_i^{t} - G^{t-1} \approx 0$ for benign clients. Hence, we rewrite Equation \ref{replacement} as follows

\begin{equation}
    G^{t} \approx G^{t-1} + \frac{\eta}{n} \frac{n}{\eta}(L_{adv}^{t} - G^{t-1}) = L_{adv}^{t},
\end{equation}
which replaces the global model with the model updates of the adversarial clients. \eugenio{If} there \eugenio{is} more than one adversarial client, the boosting factor is divided among all of them. 

\dani{Boosting techniques depends on knowing the number of clients participating in the aggregation}, which is a much more restrictive client-side knowledge condition. In practice, clients estimate this value by making several tests with different values and analysing the model \dani{updates} returned by the server. However, in the vast majority of the experimental works it is assumed the worst situation in which the adversarial clients know the number of clients of each aggregation for a better behaviour of the attack and a fair comparison between the proposed defences \cite{bib:bagdasaryan18}.

\subsubsection{Taxonomy according to the frequency}\label{sec:frequency}

\dani{As training-time phase is maintained over long periods of time}, training-time attacks can be carried out at any time of the training and on one or several occasions \cite{bib:bagdasaryan18}. We differentiate between the following two categories:

\begin{itemize}
    \item \textit{One-shot attack.} The attack is carried out in a single moment of the training, in a specific learning round. In \citet{bib:bagdasaryan18} the authors experiment with backdoor attacks at different stages of convergence and conclude that converged model attacks are more effective over several learning rounds, since the learning model does not vary and the secondary task remains injected into the global model.
    \item \textit{Multiple or adaptive attack.} The attacks are carried out continuously during the training process, either during all the learning rounds or a portion of them. They are more elaborate as the attackers have to become part of the aggregation in several rounds, but \eugenio{this kind of attack} can be more effective and stealthy \cite{DBLP:journals/corr/abs-2011-02167}.
\end{itemize}

\subsection{Privacy attacks} \label{sec:inference_attacks}

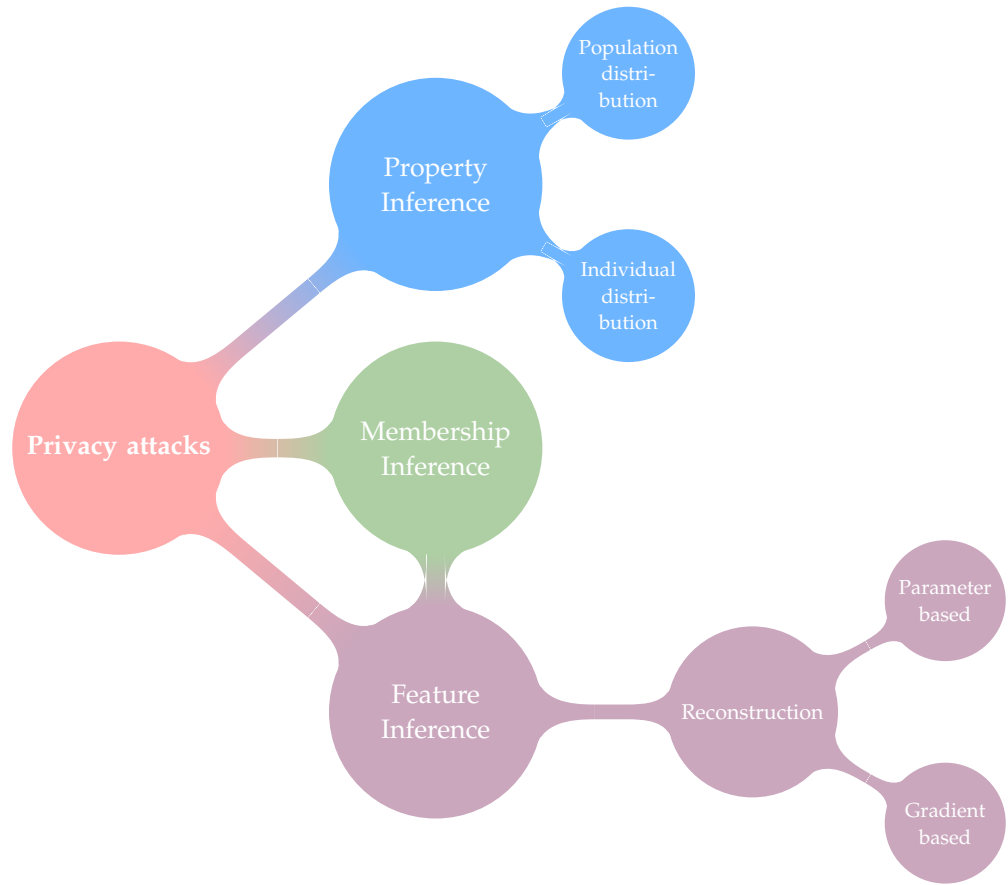
\begin{figure}[h!]
    \centering
    \input{privacy_attacks}
    \caption{Representation of the taxonomy of privacy attacks in terms of the objective of the privacy attacker.}
    \label{fig:privacy_attacks}
    \end{figure}

Privacy attacks are designed to disclose information about the participants of a machine learning task. Not only they pose a threat to the privacy of the data used to train the machine learning models, they also pose a privacy risk to those people who agreed to share their private data. FL was thought of as a privacy preserving distributed machine learning paradigm, however the learning process exposes a broad attack surface. While the private data never leaves their owner, the exchanged models are prone to memorization of the private training dataset. \dani{In this section, we present a wide taxonomy which aims to ease the understanding the diversity of privacy attacks. It is designed around the objective of the privacy attacker, a summary of it is shown in Figure \ref{fig:privacy_attacks}.}

\subsubsection{Feature inference attacks} 
\begin{figure*}[!t]
    \centering
    \includegraphics[trim={1cm 1cm 1cm 1cm},clip,width = 0.8\linewidth]{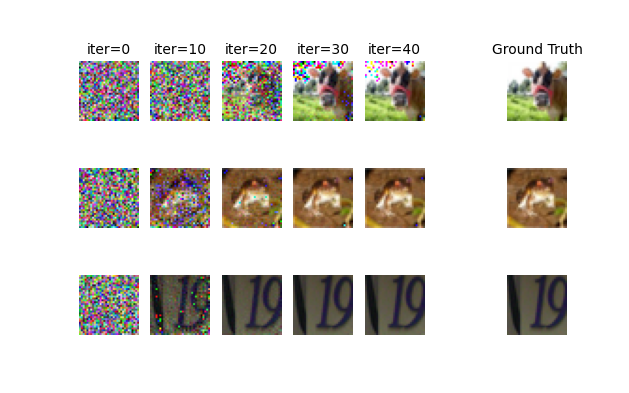}
    \caption{Gradient based Feature inference attack from \citet{Zhu2020deep} applied to CIFAR10, CIFAR100 and SVHN datasets.}
    \label{fig:reconstruct} 
\end{figure*}

Also known as \textit{Reconstruction attacks} when referring only to HFL. The aim of these attacks is recovering the dataset of a client who participates in a FL task. Usually the recovered data are images or plain text. An example of the capabilities of such attacks can be seen in Figure \ref{fig:reconstruct}. Particularly, in VFL the extracted data are the private features owned by the parties.

Accounting only for HFL, we can partition the Feature inference attacks according to the federated clients attack surface\dani{, that is, the information exchanged between the clients and the federated server}:

\begin{itemize}
    \item \textit{Gradient based}: selected clients share their gradients with the federated server in the communication rounds, that is, a federated SGD \dani{based} \dani{training procedure}. Therefore, the attack surface is the clients' gradients. To our knowledge, \citet{Zhu2020deep} are the first ones to exploit this setting. Their proposed passive attack is able to recover images and text owned by the target client. The attacker requires partial client-side knowledge, that is, accessing the gradients shared by the attacked client. However, their attack depends on its initialization and has stability issues. \citet{Zhao2020} fixes the initialization and stability problems, but the attacker requires the batch size of the clients to be 1. With the same attacker knowledge, \citet{LiHuang2019} propose a framework to measure the effectiveness of passive Feature inference attacks on logistic regression models, whose inputs are binary. \citet{Geiping2020} and  \citet{Ren2021} propose different approaches to solve the initialization and stability problems of \citep{Zhu2020deep} and their attacks can handle batches of up to 100 and 256 elements, respectively. With the same attacker knowledge, \citet{Wei2020} \dani{propose an extensive \dani{study} to measure the capabilities of passive reconstruction attacks focused on recovering images. They also propose a new attack which combines the attacks proposed in \citep{Zhu2020deep, Zhao2020}}. To our knowledge, \citet{Authors2021} are the first ones to extend and improve the attack proposed by \citet{Zhu2020deep} to a VFL setting, having the attacker third party-side knowledge. In such setting, the attacker can handle batches of up to 160 elements. When it comes to their HFL setting, the attacker requires server-side knowledge. Their proposed attack seems to be slightly better than the one proposed by \citet{Geiping2020}, but further experimentation is required to confirm their superiority. \dani{The same can be applied to \citet{Ren2021}, whose comparison with others than \citet{Zhu2020deep} remains undone.} 
    
    \item\textit{Parameter based}: selected clients share their local model parameters with the federated server in the communication rounds. Therefore, the attack surface is the clients' parameters. \dani{Focused on reconstructing training images,} \citet{Hitaj2017} presents a GAN-based active attack, where the key to train the GAN is using the global model as discriminator. The attacker requires client-side knowledge as well as extra client-side knowledge. The latter gathers the assumption that the target client and the attacker share a label, so that the inference can occur on a non-shared target label. We highlight that the attacker tricks the target client to release more information about the target label, by mislabelling the generated samples of the \dani{non-shared} target label as the shared label. In the same line, \citet{Wang2019} changes the attacker knowledge to server-side knowledge and changes the GAN architecture to a proposed multitask GAN. To further improve the effectiveness of their attack, the active attacker isolates the target client, so it does not receive global model updates. 
    
    Steeping out of GAN-based attacks, \citet{Yuan2021} focuses on reconstructing text from \eugenio{natural language processing} tasks\dani{, particularly, language modelling tasks}. The passive attacker is an observer of the federated train procedure, then she requires access to the global model at each communication round and one of the following: \eugenio{\begin{enumerate*}[label=(\arabic*)]\item to know whether the target client is selected for the communication round or \item to inject a record into the target client's training data\end{enumerate*}}. That is, she requires partial server-side knowledge and optionally partial client-side knowledge. \dani{ Their proposed attacks rely on the correlation between the privacy exposure and the clients selected in each federated aggregation step.}
\end{itemize}

The popularity of deep learning models in HFL \eugenio{cannot} be denied, however in VFL a wider range of machine learning models benefit from this setting. \citet{Luo2021} designed passive attacks for decision tree, logistic regression, random forest and neural network models. The attacker requires from the target client the feature names, types and their value range, that is, partial party-side knowledge, in addition to outsider-side knowledge. In two clients VFL setting, focusing on logistic regression and XGBoost models with party-side knowledge \citet{Weng2020} propose a passive attacker that can reconstruct the features from the other client. Although, the logistic regression attack also requires partial third party-side knowledge to gather some coefficients.


\subsubsection{Membership inference attacks} \eugenio{The main objective of these attacks is to determine whether the provided data was used to train the victim model given a client's model and some data}. In federated settings, they are commonly carried out in the \textit{model training phase}. \citet{Truex2018} study the application of Membership inference attacks to both \eugenio{non-federated} and HFL settings. In the HFL setting, their passive attack, inspired by \citet{shokri2017membership}, considers two different attacker knowledge\eugenio{: \begin{enumerate*}[label=(\arabic*)]\item where the attacker owns client-side knowledge and \item where the attacker owns outsider-side knowledge \end{enumerate*}. \citet{shokri2017membership} show that the first form of knowledge is more effective than the second one}. \citet{Nasr_2019} propose an attack with active and passive versions, each one with two options for attacker knowledge. The attack can have either client-side knowledge or server-side knowledge, where the latter is the most powerful one. Their attack consist in training a meta-classifier on the hidden layers output, the gradients, and outputs of the target client model. Such meta-classifier is a neural network with a custom architecture suited for each part of the internal state of the victim model. \dani{In the federated setting, the attack is not as effective as in the centralised scenario, so two techniques are introduced to boost the effectiveness of the attack. The first one is known as Gradient Ascent. It consists in nullifying the effect of the gradient descent on the instances used to test the attack. As a result, it broadens the difference between the data points used to train the victim model and the data points not used to train the victim model. The second one is known as Client Isolation. The objective of this technique is overfitting the victim model by not sharing with the victim client the global learning model, that is, isolating the victim client from any update.  Overfitting makes the victim model retain more information about its training dataset.} 

As data is a scarce resource, these attacks can be boosted by means of Feature inference attacks to improve the data availability \citep{Mao2019, Zhang2020, Chen2020}. \citet{Zhang2020} is a great example of using a GAN architecture for data augmentation to \eugenio{boost} the effectiveness of the passive attacker with client-side knowledge from \citet{Nasr_2019}. \dani{Increasing the attacker knowledge from client-side knowledge to client-side and server-side knowledge, and making the attacker active}, \citet{Mao2019} propose a similar use of a GAN with an attack inspired by the shadow models attack of \citet{shokri2017membership}. \citet{Chen2020} reduces the attacker knowledge to client-side knowledge and extra client-side knowledge, that is, the labels owned by each client. In addition, the attacker is passive. However, they add a new restrictive assumption, clients do not share any label.


VFL is not free from Membership inference attacks. In a two-client VFL setting, \citet{Li2021} proposes a passive attacker with party-side knowledge in a federated binary classification task.

\subsubsection{Property inference attacks} \eugenio{This kind of attacks, which are also known as \textit{attribute inference attacks}, aims at} extracting whether a property of a client or a property of the population of participants in a FL task, which might be uncorrelated with the main task of the machine learning model, is present in the FL model. In other words, the aim is to infer some property of an individual or the population which is not expected to be shared. \dani{An example of inferring an uncorrelated property is the following: consider a machine learning model whose objective is to detect faces, then the objective of the attack is inferring whether there are training images with blue-eyed faces.} As stated, we can categorise these attacks according to the target of the attacker:

\begin{itemize}
\item \textit{Population distribution}: the attacker tries to infer the distribution of a feature in a population of federated clients. In a federated SGD environment, \citet{Wang2019a} proposes a set of passive attacks. In conjunction, they can be used to infer the proportion of each label in a communication round. This attacker requires client-side knowledge and partial server-side knowledge, that is, the approximate number of clients selected by the server in a single training round, the average number of labels owned by each participant and the probable number of data samples per label. In a general HFL setting, \citet{property:Zhang2020} reduces the attacker knowledge to outsider-knowledge to perform a passive attack capable of inferring the distribution of a sensitive attribute in the training population.

\item \textit{Individual distribution}: the attacker objective is to reveal whether a target client has a property which might not be related with the main FL task. \citet{Mo2020} provide a formal framework to evaluate the property leakage of each layer of a deep learning model in a federated SGD environment. In the same federated environment, \citet{melis2019} develops both passive and active Property inference attacks, whose attacker requires only client-side knowledge. We highlight that the active attack is powered by multitask learning \citep{multitask}. Sharing the same attacker knowledge, \citet{Xu2020a} switches the environment to a standard FL setting to propose an attack with passive and active versions. The active attack employs the CycleGAN \citep{zhu2017unpaired} to reconstruct gradients with the target attribute. \citet{Chase2021} propose a Property inference attack by means of a poisoning attack. The poisoning attack requires that the attacker can modify the dataset of the target client, that is, partial client-side knowledge. Additionally, it requires the attacker to have outsider-side knowledge. In a more exotic FL environment, blockchain assisted HFL, \citet{Shen2021} propose an active attack with the requirement of server-side knowledge.

\end{itemize}

\section{Defence methods against adversarial attacks: \paco{Taxonomy}} \label{sec:defences1}



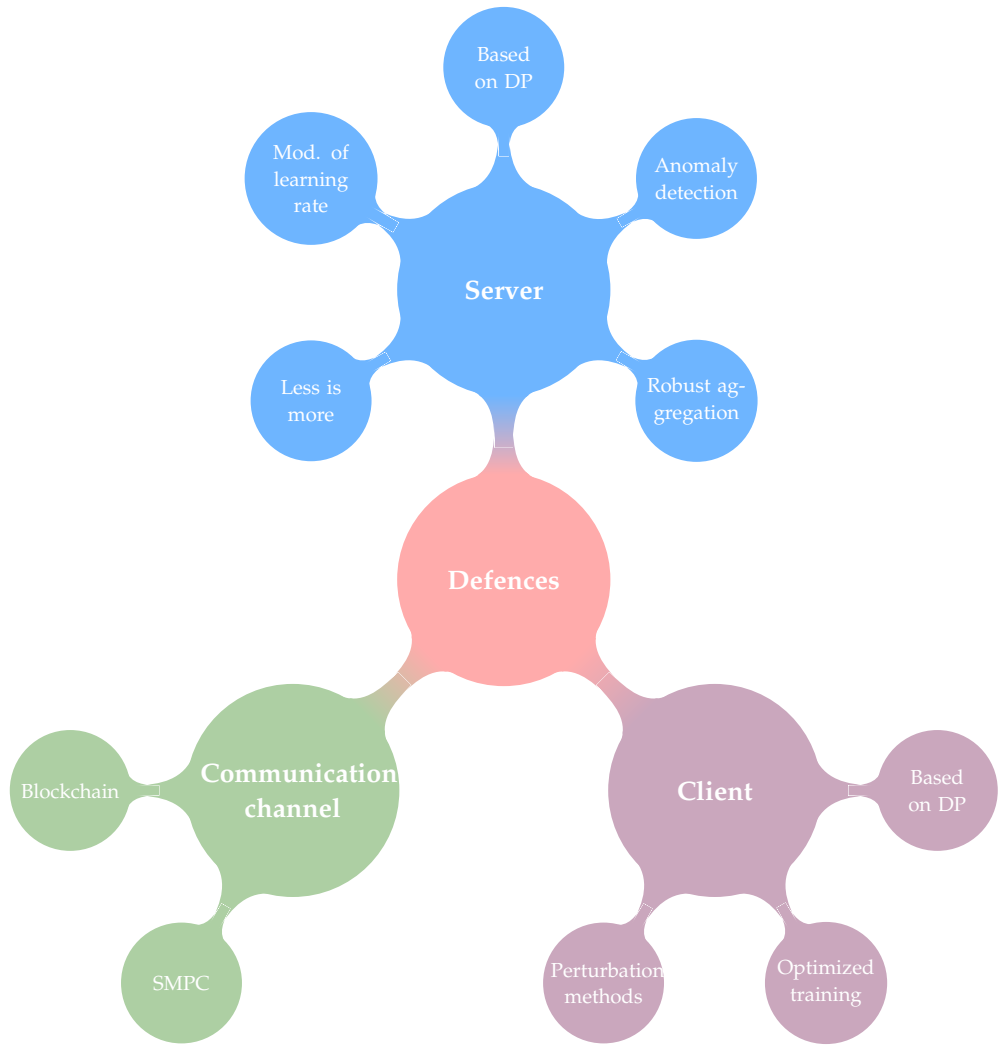
\begin{figure}[!t]
\centering
\input{defences}
\caption{Representation of the taxonomy of defences against adversarial attacks.}
\label{fig:defences}
\end{figure}

\eugenio{At the same time that the diversity and complexity of adversarial attacks against FL is enlarging,} new defences are emerging to mitigate \eugenio{their malicious effects}. \dani{While adversarial attacks can be split into disjoint categories, the same is not true for their defences as some of them are effective for more than one type of attack category. Consequently, instead of grouping defences according to the attack  defended, we categorise them }into \dani{three} main groups \dani{according to the}  federated scheme they are implemented in: the client, the server or the communication channel. \dani{Additionally, we specify for each type of defence the attacks it can defend.} In this section, we propose a taxonomy for each of these three groups of defences and highlight the most representative proposals of the state-of-the-art, which is shown in Figure \ref{fig:defences}.

\subsection{Server defences} \label{sec:server_defences}

\eugenio{The federated server is usually assumed to be reliable, because it is a controlled and accessible federated element by FL experts, in contrast to clients that are independent and inaccessible elements. Accordingly, most of defence mechanisms are implemented on the federated server.} Within this type of defences, we present the following taxonomy. \dani{Note that some defences may combine characteristics of two categories of the taxonomy. In this taxonomy, we have classified the defences according to the category that we consider best represents them.}

\subsubsection{\textbf{Robust aggregation operators}} The first and most \dani{common} approach to defend against poisoning attacks to the federated model is to use estimators that are statistically more robust than the mean to outliers or extreme values. \dani{Some aggregation operators, such as FedAvg \cite{red}, are susceptible to outliers. For that reason, many aggregation operators based on more robust estimators have been proposed.} We highlight the following ones:
\begin{itemize}
    \item \textit{Median} \cite{DBLP:journals/corr/abs-1803-01498}: It is a robust-aggregation operator based on replacing the arithmetic mean by the median of the model updates, which choose the value that represents the centre of the distribution. 
    \item \textit{Trimmed-mean} \cite{DBLP:journals/corr/abs-1803-01498}: It is a version of the arithmetic mean, consisting of filtering a fixed percentage k-\% of extreme values both below and above the data distribution. 
    \item \textit{Geometric-mean} \cite{9153949, pillutla2019robust}: It represents the central tendency or the typical value of the data distribution by using the product of their values. In other words, it chooses a reliable vector to represent the local model updates through majority voting.
    \item \textit{Norm thresholding} \cite{DBLP:journals/corr/abs-1911-07963}: It is a robust-aggregation operator, where the norm of the model updates is \dani{clipped} to a fixed value, effectively limiting the contribution of each individual update to the aggregated model.
    \item \textit{Krum and Multikrum} \cite{NIPS2017_f4b9ec30}. This aggregation operator is designed ad-hoc to prevent attacks to the federated model, so it is based on filtering out the model updates of the clients which present and extreme behaviour. For that, it sorts the clients according to the geometric distances of their model updates distributions and chooses the one closest to the majority as the aggregated model. Multikrum incorporates a $d$ parameter, which specifies the number of clients to be aggregated (the first $d$ after being sorted) resulting in the aggregated model. 
    \item \textit{Bulyan} \cite{pmlr-v80-mhamdi18a}. The authors design an \eugenio{federated} aggregation operator to prevent poisoning attacks, combining \eugenio{the} MultiKrum \eugenio{federated aggregation operator} and the trimmed-mean. Hence, it sorts the clients according to their geometric distances, and according to a $f$ parameter filters out the $2f$ clients of the tails of the sorted distribution of clients and aggregates the rest of them.
    \item \textit{Adaptive Federated Averaging (AFA)} \cite{MuozGonzlez2019ByzantineRobustFM}. Proposal of a defence mechanism against Byzantine attacks based on the weighting of each client using Hidden Markov model by means of the cosine similarity to measure the quality of model updates during training. The authors report that it discards both poor and malicious clients, improving the computational and communication efficiency.
    \item \textit{Residual-based Reweighting} \cite{fu2021attackresistant}. They propose an improvement of the median-based aggregation operator combining repeated median regression with the reweighting scheme in Iteratively Reweighted Least Squares (IRLS) based on reweighting each parameter by its vertical distance (residual) to a robust regression line.
    
    \item \textit{Sageflow} \cite{Sageflow}. A defence based on staleness-aware grouping with entropy-based filtering and loss-weighted averaging, to handle both stragglers and adversaries simultaneously. They establish a theoretical bound to provide key insights into its convergence behaviour.
    
    \item \textit{Game-theory approach} \cite{t_juegos}. The authors design the aggregation process with a mixed-strategy game played between the server and each client, where the valid actions of each client are to send good or bad model updates while the server can accept or ignore them. They weight the contribution of each client by means of the probability of providing good updates, determined employing the Nash Equilibrium property \cite{Nash48}. The main limitation is that it works only on IID training data distributions, which is unusual for real-world federated data.

\end{itemize}

\subsubsection{\textbf{Anomaly detection}} \eugenio{These defence methods consist in identifying adversarial } clients as anomalous data in the distribution of local model updates and remove them from \eugenio{the} aggregation. For this purpose, \eugenio{multivariate or adaptations of univariate anomaly detection machine learning techniques are applied.}

In \citet{Shen2016AurorDA}, the authors propose \textit{AUROR}, a defence mechanism against poisoning attacks in collaborative learning based on K-Means with $k=2$\eugenio{,} thus distinguishing between benign and suspicious clusters. Although it was a promising proposal, the main problem is that in the presence of a non-IID distribution of data between clients it could fail to identify clusters. In \citet{DBLP:journals/corr/abs-2011-02167}, they experiment with different anomaly detection mechanisms and combine the results with adaptive clipping and noise. Along the same lines, in \citet{9054676} the authors propose to divide the model updates into clusters according to the cosine distance and \citet{app8122663} proposed an incremental defence based on unsupervised deep learning anomaly detection system integrated in a blockchain process. In a similar vein, \citet{HEI2020102033} proposed an alert filter identification module in the blockchain FL process. Also in a blockchain domain, HoldOut SGD is proposed in \cite{azulay2020holdout}, which uses the holdout estimation technique in order to select the model updates that are likely to be adversarial ones. It consists in selecting two groups of clients: (1) the ones that use their private data to training in order to send their model updates and (2) a voting committee that use their private data as holdout data \eugenio{for selecting} the best model update proposals using a voting scheme. This Graph-based anomaly detection has also been proposed in \cite{8975792}, where the authors propose Sniper, a defence mechanism built upon the graph whose vertices are the updates of the local models  and the edges exists only if the two vertices are close enough. They finally identify benign local models by solving a maximum clique problem in this graph. Another example is \citet{DBLP:journals/corr/abs-1804-07474}, where the authors propose an anomaly based system based on a Gated Recurrent Unit (GRU) and test it on Internet of Things (IoT) specific databases. Along the lines of using deep learning to detect anomalies, \citet{Zhao2019PDGANAN} employ GANs by using partial classes data to reconstruct the prototypical samples of client’ training data for auditing the accuracy of each client’s model.

The main problem with anomaly-based approaches is that the model updates are likely to be very high dimensional, coming from neural networks in most cases. In \citet{DBLP:journals/corr/abs-2007-08432}, they propose to apply Principal Components Analysis (PCA) for dimensionality reduction before anomaly detection. In \citet{Li2020LearningTD} they also propose a spectral anomaly detection, which detects abnormal model updates based on their low-dimensional embeddings. The main idea is to embed both original and poisoned samples into a low-dimensional latent space and find these that differs significantly. Although these approaches reduces the problem to a low-dimensional problem, they have the limitation of losing information during the dimensionality reduction. 



\subsubsection{\textbf{Based on Differential Privacy}} Even though privacy is a topic out of the scope of adversarial attacks\dani{ to the federated model}, DP has been proven to be a viable defence method against these attacks \cite{naseri2020toward, DBLP:journals/corr/abs-1911-07963}. However, \eugenio{it is also known that DP} greatly deceives the performance of the model under circumstances of data imbalance \citep{bagdasaryan2019differential, kairouz2019advances}, which is expected to happen in most federated scenarios. Applying DP to the aggregation operator overcomes it to some extent. DP-FedAvg \citep{dp_fed_avg}, also known as Central DP, is a differentially private aggregation operator which stems from the FedAvg operator. It shares some ideas with the robust-aggregation operators, given that it removes extreme values by clipping the norm of the model updates, like the Norm thresholding operator, and then adds Gaussian noise calibrated to the clip. To provide guarantees of $(\epsilon, \delta)$-DP, the order of Gaussian noise required is high enough to reduce significantly the accuracy of the federated task. In \citet{DBLP:journals/corr/abs-1911-07963}, they introduce an alternative to Central DP aggregation operator, known as Weak DP, which shares the same aggregation procedure, but it does not guarantee $(\epsilon, \delta)$-DP nor any known privacy preserving property. It adds sufficient Gaussian noise to defeat the adversarial attack and preserve the accuracy of the federated task.

\subsubsection{\textbf{Modification of the learning rate}} One of the advantages of the server is that it sets the learning rate that controls the weighting between the previous version of the global model and the aggregate of the client model updates by means of 

\begin{equation}\label{eq:1}
    G^t = G^{t-1} + \eta \Delta(L^t_1, \dots, L^t_n)
\end{equation}

where $G^t$ is the global model in the learning round $t$, $\eta$ is the learning rate, $\Delta$ the aggregation operator and $L^t_i$ the  model update of the client $i$ in the learning round $t$. It can also decompose $\eta$ in a vector of learning rates, one per dimension. Thus, the server controls the participation in each dimension of the model updates. This decomposition approach has been used in the literature as a defence mechanism against adversarial attacks to the federated model. 

\citet{ozdayi2020defending} propose \textit{Robust Learning Rate} (RLR) as an improvement of  \textit{signSGD} \cite{DBLP:journals/corr/abs-1802-04434}. It \dani{is a} defence based on adjusting the server's learning rate $\eta$, per dimension, at each learning round according to the sign information of the clients model updates. For each dimension, they examine whether the clients agree on the direction of the model update \dani{using} a predefined threshold. If the agreement is higher than required by the threshold, the learning rate is maintained, otherwise the sign of the learning rate is changed. It can also be combined with other defences, such as those based on DP.

\subsubsection{\textbf{Less is more}} Another defence approach in the literature against adversarial attacks to the federated model is based on the fact that original task knowledge will be located in most of the weights in the model, while the weights affected by poisoning attacks will be a small portion of them. Based on this assumption, a post-training defence is proposed in \cite{DBLP:journals/corr/abs-2011-01767}, which consists of pruning the resulting global model in order to protect it against attacks that may have taken place during training. Specifically, the authors design a federated pruning method to remove redundant neurons from the neural network and to adjust the outliers of the model. They propose two pruning approaches based on majority vote and ranking vote. The main limitation is that it is usually necessary to perform fine-tuning afterwards on a validation set to compensate for the loss of accuracy caused by pruning.

In \cite{portnoy2021federated}, the authors highlight that previous works ignore the issue of unbalanced data or assume that the server owns this information. They focus on this issue and propose a practical weight-truncation-based preprocessing method\eugenio{,} which achieves quite a balance between model performance and Byzantine robustness. The novel truncation process is based on an element-wise truncation in function of some pre-fixed parameters. Although the choice of parameters is a disadvantage, the authors propose procedures for selecting them.

\subsection{Client defences}

Server defences assume that the federated server is trusted as a data collector and aggreg ator. However, this assumption might be too strong, \dani{therefore there is a requirement for defences when  the assumption of a trusted server is removed}. In such situation, defences at client level must be deployed and as a consequence, at least a portion of the clients is supposed to be benign. In contrast to server-side protection which protects clients as a whole, client-side defences are thought to be strongest as they provide protection for each client individually.


\subsubsection{\textbf{Based on Differential Privacy}} Generally, these defences are designed to defend against server-side privacy attacks, although some may prevent clients from adversarial attacks. Local DP \citep{dp_fed_avg} based on the DP-SGD algorithm presented in \citet{Abadi2016}, is the main client-side defence based on DP. \dani{Subsequently authors have proposed improvements to Local DP} in terms of DP relaxations, such as the f-DP \citep{zheng2021federated}. \citet{bu2020deep} applies f-DP to a HFL setting, achieving a better privacy analysis than \citet{Abadi2016}, that is, \dani{it provides a tighter usage of the privacy budget}. Its effectiveness against adversarial attacks have been studied \citep{naseri2020toward}, and in \citet{bib:bagdasaryan18} the reduction in performance of this technique has been related to the reduction of the effectiveness of the adversarial attack. Moreover, \citet{cao2021data} designed a successful adversarial attack aimed at Local DP protocols for frequency estimation and heavy hitter identification. In order to stop the gradient leakage, that is, privacy attacks in federated SGD settings, \citet{yadav2020differential}, \citet{hao2019towards} and \citet{wei2021gradient} made the shared gradients differentially private to protect them. \dani{If instead of exchanging parameters or gradients in HFL, clients share predictions of unlabelled data, it is possible to apply DP to protect from privacy attacks. Such setting is known as Knowledge Transfer model \citep{papernot2016semi}, and it provides privacy with a great preservation of utility using voting based approaches \citep{papernot2018scalable, zhu2020private, zhu2020voting}.}

\eugenio{Regarding defences against privacy attacks based on DP in VFL}, \dani{ \citet{wang2020hybrid} propose to perturbate the intermediate outputs shared between parties in the \textit{model training phase} of a Generalized Linear Model. Additionally, such perturbation removes the requirement of a learning coordinator and the necessity of costly Homomorphic Encryption schemes, as they are already private. However, it is a field to be explored in more depth because, to our knowledge, it is the only publication inside it.}

\citet{bhowmick2018protection} step out of the standard Local DP protocol, to relax it and provide only defence against Feature inference attacks, that is, they assume that the attacker does not have any background data about her victim.

\subsubsection{\textbf{Perturbation methods}} They are an alternative approach to provide defences against privacy attacks that are not based on DP. Its main aim is to introduce noise to \dani{the most vulnerable components of the federated model}, such as shared model parameters or the local dataset of each client, to reduce the amount of information an attacker can extract. \citet{Zhu2020deep} not only propose a Feature inference attack, they also propose some defences against it, such as gradient compression, which prunes gradients which are below a threshold magnitude. \citet{lee2021digestive} perturb the local client data with a multitask-based neural network. It preprocesses the data to increase the distance with the original data while preserving useful features for \dani{the \textit{model training phase}}.

In the same line of multitask based defences, \citet{fan2020rethinking} perturb the local training by means of a special loss in conjunction with an additional hidden neural network. \citet{sun2020provable} perturb only the parameters related to fully connected layers as they build a reconstruction procedure that can effectively reconstruct data from such layers. \citet{zhang2021matrix} propose to use the technique known as Random Sketching \citep{Woodruff} applied to shared client's parameters to defend against client-side privacy attacks. Trying to protect from the same type of client-side attacks, \citet{yang2021accuracylossless} add a kind of perturbation to the parameters that can be removed by the server, so attackers that intercepts them are not able to recover information. 



\subsubsection{\textbf{Optimised training}} \eugenio{The optimisation of the benign clients training}  may be one way to prevent the federated system from adversarial attacks. \citet{Chen2020BackdoorAO} propose to perform fine-tuning in benign clients in order to increase the impact of these clients in the aggregation. They decide which clients are benign ones by means of ``matching networks'', which consist of measuring the similarity between some inputs (the model updates) and a support set (the last global model). This way, they succeed in identifying allegedly benign clients and can conduct fine-tuning. In their experimental study, they succeed at filtering out backdoor tasks at the cost of reducing the performance of the original task. 

One of the most recent works in this line presents the client-based defence named \textit{White Blood Cell for Federated Learning} (FL-WBC) \cite{DBLP:journals/corr/abs-2110-13864}, which aims to mitigate model poisoning attacks that have already poisoned the global model. The author based the proposal on identifying the parameter space where long-lasting attacks effect on parameters resides and perturb that space during the local training of each client.

The most widespread \eugenio{training} approach aimed at preventing adversarial attacks to the federated model is \textit{adversarial training}. These defences consist in taking advantage of the robustness obtained from adversarial training in an FL setting. For example, in \cite{10.1007/978-981-15-9739-8_7} the authors propose to use pivotal training, which enables a learning model to pivot on the sensitive attributes with the aim of making the predictions independent of the sensitive attributes embedded in the training data.

\subsection{Communication channel defences}

These defences cover the space of secure implementations of FL. They enable multiple clients or parties to perform a global task, assuming the presence of some malicious actors that try to deter it. For our purposes, such actors can be embodied as the attackers that perform some adversarial attacks mentioned before. While the privacy of inputs of the global computing task is preserved, the output is revealed to some parties, if not all. Therefore, the privacy of the output is not assured, although some privacy attacks are stopped because the attacker loses access to the intermediate outputs of the global task such as the parameters or gradients shared by the clients. In other words, these defences are capable of reducing server-side knowledge to partial server-side knowledge, given that the server can only access the aggregated model or the aggregated gradients.

\paragraph{\textbf{Secure Multi-Party Computation}}

Secure Multi-Party Computation (SMPC) protocols are tightly related to Secure FL (SFL) protocols \citep{Zhu2020on}. Note that We refer to SFL protocols as FL protocols that attains the security in the simulation-based framework used to formalize the notion of security \citep{lindell2017simulate, Goldreich2001the, Goldreich2004the}. SMPC rely on Homomorphic Encryption (HE) as a key component to provide security. Consequently, HE can be regarded as the building blocks of any SMPC protocol. It provides multiple cryptographic primitives which allows for secure computations such as Secret Sharing \citep{beimel2011secret}, Zero Knowledge Proofs \citep{goldreich1994definitions} and Garbled Circuits \citep{bellare2012foundations}. Most HE based protocols only support single key encryption, which might pose a risk if the key is compromised, that is, a single point of failure. This situation has been addressed in \citep{ma2021privacy, jiang2021secure}, where the authors have developed SFL systems with multiple encryption keys.

VFL settings heavily rely on SMPC protocols to perform at the beginning of the training the private entity alignment. Additionally, when training and performing inference, partial updates and predictions are shared and the final update and prediction is computed by means of SMPC protocols.

The complexity of SMPC grows with the number of parties involved in the computation. This fact reduces the feasibility of SFL as the number of parties in a FL task can be huge \citep{kairouz2019advances}. As a consequence, the idea of full-fledged SMPC protocols that involve the entire federated training procedure are abandoned in favour of SMPC protocols that involve the communication steps in FL. As a remarkable example, a key step in HFL protocols, where SMPC protocols can ensure security and efficiency, is the aggregation step. \citet{bonawitz2017practical} defined an efficient and robust SMPC protocol for the aggregation procedure and, later on, studied its parameter selection \citep{bonawitz2019federated}. Similar ideas  and improvements have been explored by multiple authors \citep{meng2020fedmonn, kadhe2020fastsecagg, sandholm2021safe}.

To provide complete protection for both, adversarial and privacy attacks, some additional protection such as DP must be provided. SFL protocols which include DP as an additional security measure have been developed \citep{truex2019hybrid, asad2020fedopt, yong2021priv, le2021fedxgboost}. In addition, secure aggregation schemes have been improved in terms of privacy with the addition of DP mechanisms \citep{li2020secure, sabater2020distributed, ghazi2019scalable, kairouz2021distributed} .


\paragraph{\textbf{Blockchain based FL}} In  contrast to SFL protocols, Blockchain based FL enables a decentralized FL environment without single point of failure risks and improved scalability \citep{weng2019deepchain,nguyen2021federated}. However, this emerging approach inherits the already existing security issues of the blockchain: 51\% attacks \citep{li2020survey}, forking attacks \citep{wang2019corking}, double spending and reentrancy attacks on smart contract \citep{zhang2020mitigations} amongst others. In addition, it requires a way to encourage users to join the federated tasks to compensate the storage and computational usage \citep{qin2018economic}.

\section{Experimental study}\label{sec:experimental_study}


The aim of the experimental study is to analyse how attacks behave under certain circumstances and which defences are effective for which attacks, in a comparative way. For this purpose, we choose the highest-impact attack of each kind,\footnote{\eugenio{The implementation of the adversarial attacks considered in the experimental study is the provided by the authors in some cases, and the one developed by the authors of this paper thoroughly following the description of the attack on its corresponding paper.}} \eugenio{according to the previous taxonomies}, and we set the same experimental framework for each attack and test the performance of the defences in this framework. 

For each attack, we test the effectiveness of the defences in three different classification images datasets:
\begin{itemize}
    
    \item EMNIST \dani{Digits} (Extended MNIST \citep{mnist})\footnote{\url{https://www.nist.gov/itl/products-and-services/emnist-dataset}} \citep{emnist}: it is an extension of the handwritten digits dataset, MNIST. It has approximately 400,000 samples, of which 344,307 are training samples and 58,646 are test samples.
    
    
    
    
    \item The Fashion MNIST\footnote{\url{https://github.com/zalandoresearch/fashion-mnist}} \cite{DBLP:journals/corr/abs-1708-07747}: \dani{it contains a balanced set of the 10 different classes of images of clothes}, containing 7,000 samples of each class. The dataset thus consists of 70,000 samples, of which 60,000 are training samples and 10,000 test samples. 
    \item The CIFAR-10\footnote{\url{https://www.cs.toronto.edu/~kriz/cifar.html}} dataset is a labelled subset of the 80 million tiny images dataset \cite{4531741}. It consists of 60,000 32x32 colour images in 10 classes, with 6,000 images per class. \dani{The classes are: airplace, automobile, bird, cat, deer, dog, frog, horse, ship and truck}. There are 50,000 training images and 10,000 test images, which correspond to 1,000 images of each class. 
\end{itemize}

\begin{figure}
    \centering
    \includegraphics[width=13cm]{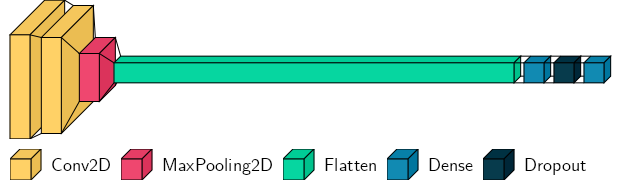}
    \caption{Convolutional network architecture used in the experimental \dani{study} for processing the EMNIST and Fashion-MNIST datasets.}
    \label{fig:conv}
\end{figure}

 \dani{For EMNIST and Fashion-MNIST we employ a standard convolutional network used in \citet{DBLP:journals/corr/abs-1911-07963} depicted in Figure \ref{fig:conv}: two convolutional layers with 3x3 kernel of 32 and 64 units followed by a 2x2 max pooling layer and a fully connected layer with 128 units with a dropout of 0.5. For the CIFAR-10 dataset, we employ a Transfer Learning approach using an EfficientNetB0 \citep{efficientnet} model pretrained on ImageNet. A fully connected layer with 256 units is added to the pretrained model. 
 }

In the following sections, we analyse the results obtained \eugenio{in the adversarial attacks to the model in Section \ref{ss_exp_at_fedmodel} and to the privacy model in Section \ref{ss_exp_at_priv}.}

\subsection{Adversarial attacks to the federated model}
\label{ss_exp_at_fedmodel}


Although the taxonomy of attacks on the model presented is broad, in this study we analyse those \eugenio{ones most used in the literature.} We assume that all the attacks are performed at training time and are multiple and static attacks, that is, the same attack is repeated in each round of learning. 

For the whole experimentation of adversarial attacks to the federated model, we consider the following \eugenio{federated distribution} of the datasets:

\begin{itemize}
    \item The federated version of the Digits dataset of EMNIST, \textit{Digits FEMNIST}. The Digits dataset of the federated version of EMNIST, where each client corresponds to an original writer.
    \item In Fashion MNIST, we set the number of clients to 500 and distribute the training data among them following a non-i.i.d distribution \dani{caused by the fact that each client randomly knows a subset of the total number of labels in the set.}
    \item In CIFAR-10, we set the number of clients to 100 and distribute the training data among them following a non-i.i.d distribution \dani{caused by the fact that each client randomly knows a subset of the total number of labels in the set.}
\end{itemize}

\dani{For all the experiments carried out in this section, we use the accuracy as \dani{evaluation} measure.}

Among the taxonomies presented, the one based on \eugenio{the existence of an specific target objective} is probably the most significant. We use this classification to divide this section into the following two subsections, corresponding to untargeted (see Section \ref{sec:untargeted}) and targeted attacks (see Section \ref{sec:targeted}).

\subsubsection{Experimental \dani{study} of untargeted attacks}\label{sec:untargeted}

Within this kind of attacks, we differentiate between: (1) those attacks that modify clients' training data, producing an alteration of the models (data-poisoning attacks) and (2) those that directly modify the weights of the learning models (model-poisoning attacks). In order to provide a variety of experimentation, we choose the following attacks:
\begin{itemize}
    \item Data-poisoning attacks: Random label-flipping attack and Out-of-distribution attack (see Section \ref{sec:poisoned}). Clearly, to make these attacks effective, we combine them with model-replacement techniques.
    \item Model-poisoning attacks: Random weights (see Section \ref{sec:poisoned}), which we also combine with model-replacement.
\end{itemize}

Regarding the ratio of adversarial clients, we considered different distributions in order to analyse the influence on both the performance of the attack and the defences. In particular, we name $x$-out-of-$n$ the situation where $x$ of the $n$ clients participating in the aggregation are adversarial ones.

We chose as defences those that have been shown to be state of the art in the literature. In particular, we use the following ones (see Section \ref{sec:server_defences}):
\begin{itemize}
    \item Median and Trimmed-mean \cite{DBLP:journals/corr/abs-1803-01498}.
    
    \item Krum and Multi-Krum \cite{NIPS2017_f4b9ec30} with \dani{different values for the parameter $d$, which detail the number of client selected. We consider} $d = 5$ and $d=20$.
    
    \item Bulyan \cite{pmlr-v80-mhamdi18a} \dani{different values for the parameter} $f$, which determines the tails of the distribution to be filtered. We consider $f=1$ and $f=2$.
\end{itemize}

In Tables \ref{tab:labelflipping}, \ref{tab:ood} and \ref{tab:random-weights} we show the results of \eugenio{assessing} the different defences in label-flipping, out-of-distribution data-poisoning attacks and random weights model-poisoning attack, respectively. In the following, we analyse the behaviour of both attacks and defences in each situation from different \paco{effectiveness and behaviour of the defences}.






\begin{table*}[!tb]
\begin{center}
\resizebox{\textwidth}{!}{\begin{tabular}{lrrrrrrrrr}
\toprule
 & \multicolumn{3}{c}{\textbf{Federated EMNIST}} & \multicolumn{3}{c}{\textbf{Fashion MNIST}} & \multicolumn{3}{c}{\textbf{CIFAR-10}}\\
\cmidrule{2-10}
 & \textbf{1-out-of-30} & \textbf{5-out-of-30} & \textbf{10-out-of-50} & \textbf{1-out-of-30} & \textbf{5-out-of-30} & \textbf{10-out-of-50} & \textbf{1-out-of-30} & \textbf{5-out-of-30} & \textbf{10-out-of-50}\\

\midrule
\textbf{No attack} & \textbf{0.965} & \textbf{0.965} & \textbf{0.962} & 0.871 & 0.871 &  0.869 & 0.835 & 0.835 & 0.823\\
\midrule

\textbf{FedAvg} & 0.159 & 0.421 & 0.400 & 0.191 & 0.366 &  0.432 & 0.118 & 0.143 & 0.244 \\
\midrule

\textbf{Trim.-mean} & 0.942 & 0.873 & 0.837 & 0.867 & 0.832 &  0.861 & 0.823 & 0.734 & 0.822 \\

\textbf{Median} & 0.931 & 0.916 & 0.909 & 0.867 & 0.847 & 0.858 & 0.828 & 0.809 & 0.828\\

\midrule
\textbf{Krum} & 0.891 & 0.870 & 0.863 & 0.726 & 0.719 & 0.747 & 0.747 & 0.761 & 0.769 \\

\textbf{MultiKrum (5)} & 0.913 & 0.927 & 0.918 & 0.840 & 0.843 &  0.825 & 0.816 & 0.823 & 0.811 \\

\textbf{MultiKrum (20)} & \textbf{0.956} &\textbf{ 0.957} & 0.950  & \textbf{0.872} & \textbf{0.872} &  0.868 & 0.843 & \textbf{0.847} & 0.851 \\

\midrule
\textbf{Bulyan (f=1)} & 0.952 & 0.781 & 0.580 & 0.868 & 0.783 &  0.787 & 0.826 & 0.659 & 0.645 \\

\textbf{Bulyan (f=5)} & 0.936 & 0.942 &\textbf{ 0.951}  & 0.861 & 0.865 & \textbf{0.872} & \textbf{0.849} & 0.845 & \textbf{0.854}  \\
\bottomrule
\end{tabular}}
\end{center}
\caption{Mean results for the \emph{label-flipping Byzantine data-poisoning attack} in terms of accuracy. We also show, in the first row, the expected accuracy with \textit{FedAvg} but without any attack.}
\label{tab:labelflipping}
\end{table*}

\begin{table*}[!tb]
\begin{center}
\resizebox{\textwidth}{!}{\begin{tabular}{lrrrrrrrrr}
\toprule
 & \multicolumn{3}{c}{\textbf{Federated EMNIST}} & \multicolumn{3}{c}{\textbf{Fashion MNIST}} & \multicolumn{3}{c}{\textbf{CIFAR-10}}\\
 \cmidrule{2-10}
& \textbf{1-out-of-30} & \textbf{5-out-of-30} & \textbf{10-out-of-50} & \textbf{1-out-of-30} & \textbf{5-out-of-30} & \textbf{10-out-of-50} & \textbf{1-out-of-30} & \textbf{5-out-of-30} & \textbf{10-out-of-50}\\
\midrule
\textbf{No attack} &\textbf{ 0.965} & \textbf{0.965} & \textbf{0.962} & 0.871 & 0.871 & 0.869  & 0.835 & 0.835 & 0.823\\
\midrule

\textbf{FedAvg} & 0.409 & 0.440 & 0.435 & 0.204 & 0.366 & 0.465 & 0.146 & 0.192 & 0.341 \\
\midrule

\textbf{Trim.-mean} & 0.945 & 0.860 & 0.853 & 0.865 & 0.834 & 0.831 & 0.820 & 0.744 & 0.740 \\

\textbf{Median} & 0.934 & 0.920 & 0.914 & 0.866 & 0.846 & 0.845 & 0.822 & 0.801 & 0.807\\

\midrule
\textbf{Krum} & 0.869 & 0.866 & 0.862 & 0.736 & 0.706 & 0.728 & 0.720 & 0.731 & 0.740 \\

\textbf{MultiKrum (5)} & 0.916 & 0.933 & 0.919 & 0.849 & 0.843 & 0.834 & 0.830 & 0.819 & 0.802 \\

\textbf{MultiKrum (20)} & \textbf{0.954} & \textbf{0.954} & \textbf{0.950} &\textbf{ 0.874} & \textbf{0.871} & 0.873 & \textbf{0.860} & \textbf{0.851} & \textbf{0.852} \\

\midrule
\textbf{Bulyan (f=1)} & 0.950 & 0.787 & 0.581 & 0.870 & 0.760 & 0.693 & 0.831 & 0.686 & 0.555 \\

\textbf{Bulyan (f=5)} & 0.935 & 0.938 & \textbf{0.950} & 0.871 & 0.865 & \textbf{0.875 }& 0.844 & 0.849 & 0.848  \\
\bottomrule
\end{tabular}}
\end{center}
\caption{Mean results for the \emph{out-of-distribution Byzantine data-poisoning attack} in terms of accuracy. We also show, in the first row, the expected accuracy with \textit{FedAvg} but without any attack.}
\label{tab:ood}
\end{table*}

\begin{table*}[!tb]
\begin{center}
\resizebox{\textwidth}{!}{\begin{tabular}{lrrrrrrrrr}
\toprule
 & \multicolumn{3}{c}{\textbf{Federated EMNIST}} & \multicolumn{3}{c}{\textbf{Fashion MNIST}} & \multicolumn{3}{c}{\textbf{CIFAR-10}}\\
\cmidrule{2-10}
 & \textbf{1-out-of-30} & \textbf{5-out-of-30} & \textbf{10-out-of-50} & \textbf{1-out-of-30} & \textbf{5-out-of-30} & \textbf{10-out-of-50} & \textbf{1-out-of-30} & \textbf{5-out-of-30} & \textbf{10-out-of-50}\\

\midrule
\textbf{No attack} & \textbf{0.965} & \textbf{0.965} & \textbf{0.962} & 0.871 & 0.871 & 0.869  & 0.835 & 0.835 & 0.823\\
\midrule

\textbf{FedAvg} & 0.099 & 0.099 & 0.100 & 0.100 & 0.101 & 0.099 & 0.099 & 0.099 & 0.100  \\
\midrule

\textbf{Trim.-mean} & 0.953 & 0.103 & 0.099 & 0.875 & 0.100 & 0.099 & 0.860 & 0.099 & 0.099  \\

\textbf{Median} & 0.936 & 0.935 & 0.934 & 0.865 & 0.861 & 0.855 & 0.849 & 0.866 & 0.864 \\

\midrule
\textbf{Krum} & 0.831 & 0.865 & 0.854 & 0.715 & 0.745 & 0.734 & 0.718 & 0.716 & 0.799 \\

\textbf{MultiKrum (5)} & 0.932 & 0.922 & 0.919 & 0.834 & 0.834 & 0.827 & 0.816 & 0.811 & 0.816  \\

\textbf{MultiKrum (20)} & \textbf{0.956} & \textbf{0.957} & 0.951 & \textbf{0.876} & \textbf{0.875} & 0.867 & 0.848 & \textbf{0.848} & \textbf{0.853} \\

\midrule
\textbf{Bulyan (f=1)} & 0.959 & 0.099 & 0.099 & 0.099 & 0.100 & 0.099 & 0.852 & 0.099 & 0.099  \\

\textbf{Bulyan (f=5)} & 0.937 & 0.937 & \textbf{0.951} & 0.874 & 0.869 & \textbf{0.874} & \textbf{0.850} & 0.841 & 0.851  \\
\bottomrule
\end{tabular}}
\end{center}
\caption{Mean results for the \emph{random weights Byzantine model-poisoning attack} in terms of accuracy. We also show, in the first row, the expected accuracy with \textit{FedAvg} but without any attack. }
\label{tab:random-weights}
\end{table*}


\paragraph{Effectiveness of the attack} If we compare the effectiveness of the attack in function of the type of attack, we conclude that the most damaging attack is the random weights attack. In fact, this attack \eugenio{manages to totally confuse the federated model, to the extent that it behaves as a most frequent label classification model.} If we focus on the data-poisoning attacks, we get that the label-flipping attack is sightly more effective than the out-of-distribution attack. This is probably because the \dani{label-flipping attack} learns miss-labelled samples from within the distribution, while the \dani{out-of-distribution attack}, theoretically, only adds error to samples from outside the distribution. 

Regarding the ratio of adversarial clients participating in each aggregation, we found that there are significant differences, being the most effective one carried out by a single adversarial client (1-out-of-30). While this may seem contradictory, there is an explanation. When the attack is carried out by several clients, the boosting factor is divided among these adversarial clients. This divides \eugenio{the strength of the attack among all the adversarial clients, which thus weak the power of the attack}, whereas when carried out by a single client, all the boosting is reflected in a single attacker, making it more effective.


\paragraph{Behaviour of the defences} As a general rule, the defences that best mitigate the effect of the attacks are Multikrum (20) and Bulyan (f=5), with MultiKrum (20) standing out slightly. As we have \eugenio{shown}, although Bulyan is presented as an improvement of MultiKrum in combination with trimmed-mean, if the pre-selected clients are benign clients, this truncation is not necessary and even superfluous. On the other hand, the more basic defences such as median and trimmed-mean show good enough behaviour in some experiments, even outperforming MultiKrum and Bulyan with some parameters.

This superiority of the most basic defences over MultiKrum and Bulyan with \eugenio{specific parameters values evidences} the high dependence of these defences on \eugenio{the values of the input parameters. This behaviour matches with the assertion of the authors of MultiKrum and Bulyan, they are the most robust defences with the optimal value of the input parameters. This dependency on the values of the input parameters represents an obstacle for the use of this defences, since the value of some parameters is difficult to know, like the number of adversarial clients}. A clear example of this problem is Bulyan (f=1) in the random weights Byzantine model-poisoning attack, whose results are comparable to using no defence at all by filtering out too few adversarial clients.

To conclude, untargeted attacks are highly effective, especially those based on model-poisoning, which achieve random behaviour in the federated model. The defences proposed in the literature perform reasonably well, substantially improving the effect of the attacks, even the simplest ones. However, none of them manage to completely dissipate the attack, and the best-performing ones are highly dependent on configuration parameters, so there is still \eugenio{room for improvement} in designing defences against Byzantine attacks. 

\subsubsection{Experimental \dani{study} of targeted attacks}\label{sec:targeted}

In order to make a sufficiently broad experimental study, in this section we consider backdoor attacks from the two main groups presented: (1) Input-instance-key strategies and (2) pattern-key strategies. With respect to attacks implementing input-instance-key strategies, we perform a single attack where the target samples correspond to some samples belonging to the adversarial clients for each dataset and associate them with a specific target label. However, with respect to the pattern-key attacks, we choose for each dataset a different static, single and accessory injection pattern. 

We chose the state of the art against \dani{Backdoor} attacks as baselines. In particular, we use the following ones (see Section \ref{sec:untargeted}):
\begin{itemize}[noitemsep]
    \item Median and Trimmed-mean \cite{DBLP:journals/corr/abs-1803-01498}.
    \item Norm-clipping \cite{DBLP:journals/corr/abs-1911-07963}.
    \item Weak Differential Privacy (Weak DP) \citet{DBLP:journals/corr/abs-1911-07963}.
    \item Robust Learning Rate (RLR) \citet{ozdayi2020defending}.
\end{itemize}
For these defences based on clipping and noise addition, we use $M$ and $\sigma$ to specify both the clip factor and the noise added, respectively. For the experiments, we choose the values recommended by the authors.

\paragraph{\dani{Study} of Input-instance-key attacks}

In Table \ref{input-attacks-results} we show the results obtained after testing the input-instance-key attack and the different defences. For the implementation, we randomly select some samples of the adversarial clients and associate them with the target label "0". We evaluate the effectiveness of the attack, showing both the original and backdoor performances. We measure the original performance using the mean accuracy in the original test dataset and the backdoor performance by means of the mean accuracy in the set of selected samples for the attack.

\begin{table*}[h!]
\vskip 0.15in
\begin{center}
\begin{scriptsize}
\begin{tabular}{lrrrrrrrr}
\toprule
 &     &    & \multicolumn{2}{c}{\textbf{Federated MNIST}} & \multicolumn{2}{c}{\textbf{Fashion MNIST}} & \multicolumn{2}{c}{\textbf{CIFAR-10}}\\  \cmidrule{4-9}
 & $M$ & $\sigma$ & \textbf{Original} & \textbf{Backdoor} & \textbf{Original} & \textbf{Backdoor} & \textbf{Original} & \textbf{Backdoor}\\
\toprule
\textbf{No attack} & 0 & 0 & \textbf{0.965} & - &  0.871  & - & 0.835 & - \\
\midrule
\textbf{FedAvg} & 0 & 0 & 0.866 & 0.823 & 0.804 & 0.944 & 0.612 & 0.903\\
\textbf{\nuria{Median}} & \nuria{0} & \nuria{0} & \nuria{0.944} & \nuria{0.030} & \nuria{\textbf{0.875}} & \nuria{0.032} & \nuria{0.861} & \nuria{0.140}\\
\textbf{\nuria{Trim.-mean}} & \nuria{0} & \nuria{0} & \nuria{0.952} & \nuria{0.025}& \nuria{0.872} & \nuria{0.016} & \nuria{\textbf{0.863}} & \nuria{0.133} \\
\textbf{NormClip} & 3 & 0 & \textbf{0.960} & 0.876 & 0.863 & 0.144 & 
0.843 & 0.115\\
\textbf{Weak DP} & 3 & 2.5e-3 & 0.937 & 0.157 & 0.843 & 0.119 & 0.823 & 0.093\\
\textbf{RLR} & 0.5/0.5/1 & 1e-4 & 0.954 & \textbf{0.012} & 0.863 &  \textbf{0.002} & 0.853 & \textbf{0.014}\\
\bottomrule
\end{tabular}
\end{scriptsize}
\end{center}
\caption{Mean results for the input-instance backdoor attack in terms of accuracy. We also show, in the first row, the expected accuracy with \textit{FedAvg} but without any attack.}
\label{input-attacks-results}
\vskip -0.1in
\end{table*}

If we first analyse the effectiveness of the attack (see row of \textit{FedAvg} and \textit{Backdoor} columns) we find the attack is relatively effective, with the result in Fashion MNIST standing out, and always being higher than $0.82$ \eugenio{of accuracy}. However, if we focus on the stealthiness we note that this type of attack lacks this valuable quality, even affecting the performance on the original task (see row of \textit{FedAvg} and \textit{Original} columns) in 22 points of accuracy (CIFAR-10).

Regarding the performance of the defences, we find that every one of the defences leads to a substantial improvement, both increasing the original task accuracy and reducing the backdoor task accuracy. In addition, we would like to highlight the good performance of the simpler defences, such as trimmed-mean, which achieves very competitive results. If we analyse the state-of-the-art defences (Weak DP and RLR), we found the results to be appropriate, but perhaps a mite disappointing on a complexity-performance trade-off compared to the other defences. Moreover, there are likely to be other $M$ and $\sigma$ parameters that optimize the results of these defences, but there are not known in advance, which is the main weakness of such parameter-dependent defences.

To conclude, input-instance-key backdoor attacks are considerably powerful, performing better in the backdoor task than in the original one, but being too eye-catching and detrimental to the original task. Moreover, although the defences mitigate the effect of the attack, none of them completely dissipate it, so there is still plenty of scope for further research.

\paragraph{\dani{Study} of the pattern-key attacks}

The Table \ref{pattern-attacks-results} shows the results obtained after testing the different pattern-key attacks with the considered defences. We implement the attacks by randomly selecting the adversarial clients and poisoning some of their samples with different patterns. In particular, we use the following patterns of different levels of difficulty according to the number of pixels: (1) one single black pixel for Federated MNIST, (2) a red cross of length 4 for Fashion MNIST (8-pixel pattern) and (3) a white pixel in each of the corners of the image (4-pixel pattern) for CIFAR-10. We evaluate both the effectiveness and the stealthiness of the attack. We measure the stealthiness of the attack by means of the mean accuracy obtained in the original task (Original). We also evaluate the effectiveness of the attack in terms of two \eugenio{additional} tests: (1) Backdoor, which contains the poisoned samples of the adversarial clients and (2) Test, which represents the test of the backdoor task and is composed of test samples poisoned following the specific pattern. Therefore, an attack will be more effective the higher performance it obtains in both the original and the backdoor task, while a defence will be better if it manages to maintain the performance in the original task while decreasing as much as possible the performance in the backdoor task. 

\begin{table*}[h!]
\vskip 0.15in
\begin{center}
\begin{scriptsize}
\begin{tabular}{lrrrrrrrrrrr}
\toprule
 &     &    & \multicolumn{3}{c}{\nuria{\textbf{Federated MNIST}}} & \multicolumn{3}{c}{\textbf{Fashion MNIST}} & \multicolumn{3}{c}{\textbf{CIFAR-10}}\\
 \cmidrule{4-12}
 & $M$ & $\sigma$ & \textbf{Original} & \textbf{Backdoor} & \textbf{Test} & \textbf{Original} & \textbf{Backdoor} & \textbf{Test}  & \textbf{Original}  & \textbf{Backdoor} & \textbf{Test}\\
\toprule
\textbf{No attack} & 0 & 0 & 0.965 & - & - & 0.871 & - & - & 0.835  & - & - \\
\midrule
\textbf{FedAvg} & 0 & 0 & 0.974 & 1.0 & 1.0 & 0.843 & 0.999 & 0.944 & 0.413 & 1.0 & 0.99\\
\textbf{\nuria{Median}} & 0 & 0 & 0.954 & 0.009 & 0.015 & \textbf{0.873} & 0.067 & 0.053 & 0.854 & 0.193 & 0.183\\
\textbf{\nuria{Trim.-mean}} & 0 & 0 & 0.966 & 0.011 & 0.014 & 0.872 & 0.052 & 0.065 & 0.853 & 0.194 & 0.170\\
\textbf{NormClip} & 1 & 0 & \textbf{0.968} & 0.055 & 0.053 & 0.843 & 0.143 & 0.164 & 0.834 & 0.143 &  0.131\\
\textbf{Weak DP} & 1 & 2.5e{-3}  & 0.935  & 0.093  &0.0175 & 0.869 & 0.053 & 0.074  & \textbf{0.859} & 0.144 & 0.170 \\
\textbf{RLR} & 1 & 1e{-4}  &  0.962 & \textbf{0.008} & \textbf{0.008} & 0.870 & \textbf{0.020} & \textbf{0.031} & 0.856 & \textbf{0.073} & \textbf{0.061}\\
\bottomrule
\end{tabular}
\end{scriptsize}
\end{center}
\caption{Mean results for the \emph{pattern-key backdoor attack} in terms of accuracy. We also show, in the first row, the expected accuracy with \textit{FedAvg} but without any attack. The best result for each of the test sets is highlighted in bold.}
\label{pattern-attacks-results}
\vskip -0.1in
\end{table*}


Regarding the effectiveness of the attack without any defence (see row of the \textit{FedAvg} and \textit{Backdoor} and \textit{Test} columns), \eugenio{it reaches a performance of 100\% or close to it of accuracy, which shows it harmfulness.} However, if we analyse the stealthiness of the attack (see row of the \textit{FedAvg} and \textit{Original} columns), the conclusions depend on the dataset. While in Federated MNIST and Fashion MNIST the performance in the original task is maintained or even improved, the performance in the original task in CIFAR-10 is reduced by up to half.

Regarding the behaviour of the defences, we also obtain a substantial improvement with respect to the scenario without any defence with all of them. As in the untargeted attacks, the simplest defences obtain competitive results, even outperforming the most complex defences in some situations. In general, deciding which defence is superior is not a trivial task. Since it is a matter of achieving the best trade-off between performance in the original task and dissipation of the backdoor attack. For example, RLR achieves in Federated EMNIST the best defence against \eugenio{the} attack\eugenio{, but it} is more detrimental to performance on the original task. However, in general, we can affirm that it is the best performing defence, standing out particularly in CIFAR-10.

To conclude, pattern-key backdoor attacks are highly threatening attacks, as they achieve almost 100\% success in the backdoor task, without, in most cases, harming the performance of the original task. Defences manage to dissipate the effect of the attack in the backdoor task, but in most cases impair performance in the original task. Therefore, in this case, the key is to find the trade-off between mitigating the attack and not harming the performance of the model.

\subsection{Privacy attacks}
\label{ss_exp_at_priv}

Even tough there is a wide range of privacy attacks, in this section we study those \eugenio{which} meet the following requirements: 
\begin{enumerate}
    \item The attack is performed while the federated model is being trained. As a consequence, most defences are aimed to make the training secure from privacy attacks. Alternatively, the defences mask or perturb the shared information to make it less vulnerable. 
    
    \item The description of the attack and its setup in its publication is enough to implement it or an implementation which matches the publication is publicly available. The same applies for defences.
\end{enumerate}

The found privacy attacks that matched our requirements allowed us to divide this section into the following two subsections, corresponding to Membership inference attacks (see Section \ref{exp:mia}) and Feature inference attacks (see Section \ref{exp:rec}), restricted to HFL scenarios.

\subsubsection{Experimental \dani{study} of Membership inference attacks} \label{exp:mia}

We choose to implement the federated white-box Membership inference attack from \citet{Nasr_2019} using the source code publicly available for the white-box centralized setting\footnote{https://github.com/privacytrustlab/ml\_privacy\_meter} as there is no public implementation of the federated version. Both clients and server can be the attacker. On the one hand, when the attacker is the client, her objective is to infer the membership of data points belonging to other clients. On the other hand, when the attacker is the server every client is attacked individually, thus the objective is to infer the membership of data points for each client. We mainly focus on their server side attack or global attacker as it is the most powerful, that is, it poses the highest treat to privacy.

We make our federated scenarios \dani{the same as} the ones proposed in \citet{Nasr_2019}, which represents a small population of clients with big amounts of sensitive data such as banks or hospitals, willing to jointly train a privacy preserving deep learning model. As each client owns great quantities of data, some records can be duplicated among them, that is, the dataset owned by each client is sampled uniformly with replacement from the following datasets: EMNIST, Fashion MNIST and CIFAR-10. Consequently, each of them is divided between 4 clients and each client owns a sample of half the size of the entire dataset, sampled with replacement.

Each federated task is run for 300 rounds where each client shares her local model after each local training epoch, the attacker observes the rounds: 50, 100, 200, 250 and 300. The attack is trained for 100 epochs and the model with best testing accuracy is selected. The attacker training dataset is made of 4000 random samples belonging to each attacked client, 4000 random samples which do not belong to any client and each one is labelled according to its membership to the attacked client. For all the experiments, the batch size is set to 32. We highlight that this federated setup is \dani{taken} from \citet{Nasr_2019}.

We report the averaged accuracy and AUC of the global attacker in the described settings in Table \ref{tab:mia_all}. \dani{Note that, the membership inference attack is performed by a binary classifier, therefore the choice of the classification threshold is key to separate between member and non member instances. An attacker with background knowledge may have the ability of selecting a classification threshold that maximizes the separation between member and non members, leading to a greater privacy leakage \citep{yeom}.} While the authors of the attack focus on reporting the accuracy, we have found in our experiments that the AUC metric better \eugenio{shows} the capabilities of the attacker, due to the fact that AUC is independent of the classification threshold used to perform the inference. \dani{This decision is also driven by the fact that a single classification threshold only represents a possible attacker, therefore we need a way of evaluating every possible attacker, including those with great amounts of background knowledge.} We can observe that in our experiments the attack is barely effective, as both accuracy and AUC are close to 0.5. We also highlight that the Gradient Ascent technique does not bring significant performance improvements, probably because it is hard to calibrate. While in the MNIST dataset we \eugenio{see} that the attack is not successful, in the other the membership of some instances is revealed, so there is a privacy leak, although it is very small. 


We also report the success of the attack with the state-of-the-art defence Local DP in Table \ref{tab:mia_all}\eugenio{. The privacy budget in each client of the Local DP is } $\epsilon=3, \delta=10^{-5}$, which is considered in the literature to be a high privacy budget. We employ the AutoDP framework\footnote{\url{https://github.com/yuxiangw/autodp}} to calibrate the differentially private Gaussian noise to the privacy budget using Renyi DP \citep{wang2019subsampled}. We can observe that this defence is quite successful as it avoids leaking any membership information, thus making the attack classifier behaves randomly.


\begin{table}[!htp]\centering
\scriptsize
\begin{tabular}{lrrrrrrrrr}\toprule
&\multicolumn{4}{c}{Without Local DP defence} &\multicolumn{4}{c}{With Local DP defence} \\\cmidrule{2-9}
&\multicolumn{2}{c}{Client Isolation} &\multicolumn{2}{c}{\makecell{Client Isolation +\\Gradient Ascent}} &\multicolumn{2}{c}{Client Isolation} &\multicolumn{2}{c}{\makecell{Client Isolation +\\Gradient Ascent}} \\\cmidrule{2-9}
&Accuracy &AUC &Accuracy &AUC &Accuracy &AUC &Accuracy &AUC \\\midrule
MNIST &0.501 &0.502 &0.489 &0.502 &0.500 &0.497 &0.496 &0.500 \\
Fashion MNIST &0.513 &0.546 &0.511 &0.516 &0.500 &0.499 &0.497 &0.500 \\
CIFAR-10 &0.540 &0.551 &0.500 &0.528 &0.500 &0.500 &0.500 &0.500 \\
\bottomrule
\end{tabular}
\caption{Accuracy and AUC of the global federated attack from \citet{Nasr_2019} with and without Local DP defence. }\label{tab:mia_all}
\end{table}

In Table \ref{tab:fed_task_acc}, we can see the accuracy of the federated task with the attack. As noted before, the Gradient Ascent technique degrades the accuracy of the federated task, mainly due to the fact that some of the instances which were ascent belong to the federated test set. While this is true for the MNIST and Fashion MNIST datasets, it is not true for the CIFAR-10 dataset. It might be because of the transfer learning approach used for this dataset being more resilient to gradient direction changes. As expected, DP based defences reduce the accuracy of the federated task. The smallest reduction of federated task accuracy is achieved with the CIFAR-10 dataset, which confirms that the transfer learning approach is more resilient to gradient changes, moreover the Gradient Ascent technique does not change significantly the accuracy when applied.



\begin{table}[!htp]\centering
\scriptsize
\begin{tabular}{lrrrrr}\toprule
&\multicolumn{2}{c}{Without Local DP defence} &\multicolumn{2}{c}{With Local DP defence} \\\cmidrule{2-5}
&Client Isolation &\makecell{Client Isolation +\\Gradient Ascent} &Client Isolation &\makecell{Client Isolation +\\ Gradient Ascent} \\\midrule
MNIST &0.990 &0.100 &0.672 &0.100 \\
Fashion MNIST &0.910 &0.100 &0.579 &0.100 \\
CIFAR-10 &0.862 &0.862 &0.686 &0.668 \\
\bottomrule
\end{tabular}
\caption{Federated task accuracy while the global federated attack from \citet{Nasr_2019} is performed with and without Local DP defence.}\label{tab:fed_task_acc}
\end{table}


In this experimental study, we have explored the performance of a Membership inference attack on a federated setting of few clients with big amounts of data. We have found that the success of the attack is small, even though the membership of some instances was revealed. The DP based defence stopped these leakages of privacy, at the cost of a considerable reduction of the federated task accuracy. Additionally, we have found that using a transfer learning approach might reduce the impact of DP in the federated task accuracy while also being resilient to the Gradient Ascent technique which have drastically reduced the federated task accuracy with the other datasets and deep learning approaches.

\subsubsection{Experimental study of Feature inference attacks} \label{exp:rec}

We study multiple gradient based Feature inference attacks, which use stolen gradients from the federated training procedure, particularly we focus on the attacks described in \citet{Zhu2020deep, Geiping2020, Wei2020}. In order to do it, we use the code provided with each publication, which is publicly available.\footnote{\url{https://github.com/mit-han-lab/dlg}}\textsuperscript{,}\footnote{\url{https://github.com/JonasGeiping/invertinggradients}}\textsuperscript{,}\footnote{\url{https://github.com/git-disl/CPL\_attack}}

The federated scenario which fits these attacks is the following: clients with very little data, such as IoT devices or smartphones, which run a federated task where they share gradients from small batches. We study under which circumstances we can reconstruct images from gradients. Our \dani{study} focus on three aspects to evaluate the success of these attacks: 

\begin{itemize}
\item \textbf{The success rate}\eugenio{.} The approximate probability of convergence of each attack. The majority of the attacks studied in this section are known to have stability issues, that is, their convergence greatly depends on the initialization seed used to bootstrap them. For \citet{Wei2020} and \citet{Zhu2020deep}, we choose as initialization a geometric pattern which improves both convergence rate and speed. It consists in covering a small portion of the initialization space with a random image and duplicate it to fill the feature space. In our experiments, we choose 1/4 of the feature space as in \citet{Wei2020}. For the attack of \citet{Geiping2020}, we choose random initialization, as it does not seem to be affected by the choice of the initialization pattern.

\item \textbf{The training stage of the local model at which the attack can succeed}\eugenio{.} Most of the \eugenio{studied} attacks consider an untrained model as they claim that the attack can run at any point of the training procedure, however this claim does not seem to have a lot of experimental support. As a consequence, we want to confirm such claims and find whether the stage of training of the local model is relevant to the success of the attack. 

\item \textbf{The success of the defences against Feature inference attacks}\eugenio{. We} study the performance of two state-of-the-art defences: gradient compression and the addition of Gaussian noise. Which are known to thwart the effectiveness of the attack from \citet{Zhu2020}, so we \eugenio{evaluate whether} these defences are also applicable to the other attacks.
\end{itemize}

We begin our study analysing the success rate of each attack, as they are known to suffer from stability issues \citep{Wei2020}. We run each attack with gradients from an untrained simple convolutional model \textit{LeNet} \citep{lecun1989backpropagation} as in \citep{Wei2020, Zhu2020deep} with a batch size of 1. Each attack is run until one of the following conditions is satisfied:

\begin{itemize}
\item Success condition: for the attacks \citep{Wei2020, Zhu2020deep}, we consider that the attack is \eugenio{successful} if the Mean Square Error (MSE) with respect to the target image to reconstruct is smaller than 0.5 and the Multi-Scale Structural Similarity (SSIM) \citep{wang2003multiscale} is greater than 0.5. The purpose of these criterions is twofold, the former ensures that the reconstructed image is close enough to the target image and the latter ensures that the reconstructed image is perceptibly similar to the target image. 

\item Failure condition: if the maximum number of epochs set for the attack is reached without satisfying any of the conditions stated below, then we consider the attack is marked as a failure. In other words, the attack as failed to converge.

\end{itemize}

Additionally, we want to study whether the training stage at which the attack is performed is relevant. To do so, we run each attack at different moments of the local training process: \dani{before any training,} after 1, 5 and 10 rounds of training. Each attack is going to try to reconstruct an image that belongs to their training set, but it has not been used to train the model previously. We report the success rate of each attack across 25 runs, using the same end conditions specified before.

The experimental results of the study of \textbf{the success rate} and \textbf{the training stage of the local model at which the attack can succeed} are shown in Tables \ref{tab:dlg}, \ref{tab:cpl} and \ref{tab:gei}. \eugenio{First}, we highlight the results from Table \ref{tab:gei} \eugenio{that} show that the attack from \citet{Geiping2020} is independent of the considered training stage of the local model. The same is not true for the results in Tables \ref{tab:dlg} and \ref{tab:cpl}. In its first column of results, we can see that the attacks have almost no issues to converge when the local model is not trained, so we can conclude that if the appropriate initialization method is chosen, the attacks are almost 100\% guaranteed to converge. If we observe the remaining columns of the Tables \ref{tab:dlg}, \ref{tab:cpl}, the results change considerably. The attack from \citet{Wei2020} (Table \ref{tab:cpl}) has slightly better convergence rates than the attack from \citet{Zhu2020deep} (Table \ref{tab:dlg}), both show a similar trend: the more trained is the model, the harder it is for the attacks to achieve success. 

\begin{figure}%
    \centering
    \includegraphics[trim={0 0 0 0.65cm},clip,width=2.5cm]{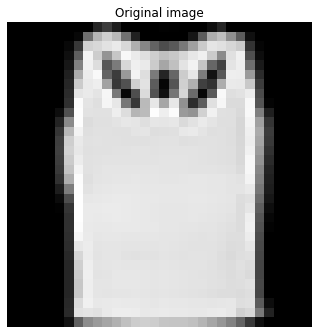} \qquad %
    \includegraphics[trim={0 0 0 0.7cm},clip,width=2.5cm]{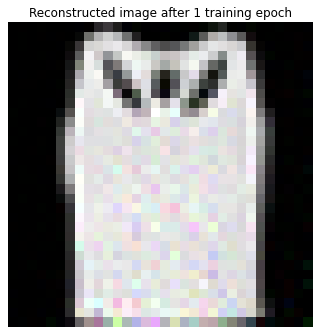}%
    \includegraphics[trim={0 0 0 0.65cm},clip,width=2.5cm]{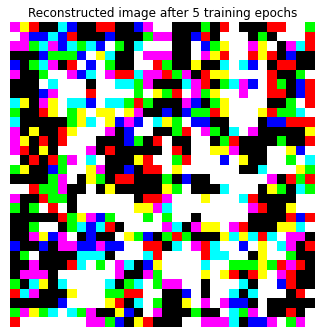}%
    \includegraphics[trim={0 0 0 0.6cm},clip,width=2.5cm]{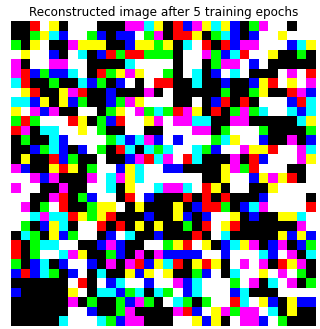}%
    
    \caption{From left to right, reconstruction using the attack of \citet{Wei2020} of an image with label 0 from Fashion-MNIST dataset \dani{which correspond to the t-shirt/top category}, after 1, 5 and 10 epochs of local training.}%
    \label{fig:fmnist-wei}%
\end{figure}

\eugenio{T}he complexity of the dataset has an important role in the success of the attacks from \citep{Wei2020} and \citet{Zhu2020deep}. Both EMNIST and Fashion-MNIST are considered easier datasets, as there are many works that achieve high training accuracy after few epochs of training \citep{WanICML2013, NIPS2015_6855456e}. The same is not true for CIFAR-10, as more complex models are required to achieve a reasonable accuracy \citep{pmlr-v139-tan21a, graham2021levit}. EMNIST and Fashion-MNIST images are hard to reconstruct after 1 training epoch, that is, the gradients after 1 training epoch leak little information about the datasets. An example of such difficulties is shown in Figure \ref{fig:fmnist-wei}. In contrast, in CIFAR-10 the training model takes longer to converge and the gradients leaks a lot of information, even after 10 epochs of training. The main reason that allow us to understand this behaviour is the fact that both attacks try to mimic the structure and content of the shared gradient (that is, minimizing the MSE between the shared gradient and the reconstructed image), so the more information is stored in the gradient, the easier is the reconstruction process. In other words, gradients that \eugenio{more significantly} change the weights of the model make the reconstruction process easier. This is not true for the attack from \citep{Geiping2020}, as its objective is to minimize the cosine similarity between gradient vectors.

\begin{table}[!htp]\centering
\scriptsize
\begin{tabular}{lrrrrr}\toprule
Dataset &Before training &After 1 training epoch &After 5 training epochs &After 10 training epochs \\\midrule
EMNIST &1 &0 &0.04 &0 \\
Fashion-MNIST &1 &0.28 &0 &0.08 \\
CIFAR-10 &0.96 &0.80 &0.60 &0.68 \\
\bottomrule
\end{tabular}
\caption{\dani{Success rate of 25 trials of reconstructing an image from a shared gradient of a local model with the attack from \citet{Zhu2020deep}. We run the attack at different stages of training of the local model. \textit{Before training} means that the local model has not been trained at all.}}
\label{tab:dlg}
\end{table}

\begin{table}[!htp]\centering
\scriptsize
\begin{tabular}{lrrrrr}\toprule
Dataset &Before training &After 1 training epoch &After 5 training epochs &After 10 training epochs \\\midrule
EMNIST &1 &0 &0 &0 \\
Fashion-MNIST &1 &0.32 &0.12 &0.24 \\
CIFAR-10 &1 &0.96 &0.84 &0.80 \\
\bottomrule
\end{tabular}
\caption{\dani{Success rate of 25 trials of reconstructing an image from a shared gradient of a local model with the attack from \citet{Wei2020}. We run the attack at different stages of training of the local model. \textit{Before training} means that the local model has not been trained at all.}} \label{tab:cpl}
\end{table}

\begin{table}[!htp]\centering
\scriptsize
\begin{tabular}{lrrrrr}\toprule
Dataset &Before training &After 1 training epoch &After 5 training epochs &After 10 training epochs \\\midrule
EMNIST &1 &1 &1 &1 \\
Fashion-MNIST &1 &1 &1 &1 \\
CIFAR-10 &1 &1 &1 &1 \\
\bottomrule
\end{tabular}
\caption{\dani{Success rate of 25 trials of reconstructing an image from a shared gradient of a local model with the attack from \citet{Geiping2020}. We run the attack at different stages of training of the local model. \textit{Before training} means that the local model has not been trained at all.}}

\label{tab:gei}
\end{table}

To end our study, we study the performance of two state-of-the-art defences:

\begin{itemize}[noitemsep]
\item Gradient compression with 20\% sparsity.
\item The addition of Gaussian noise with variance of $10^{-2}$.
\end{itemize}

We run each attack with defences 25 times with a batch size of 1, with the model untrained and report the success rate of each attack.

\begin{table}[!htp]\centering
\scriptsize
\begin{tabular}{lrrrrrrr}\toprule
&\multicolumn{2}{c}{Attack of \citet{Zhu2020deep}} &\multicolumn{2}{c}{Attack of \citet{Wei2020}} &\multicolumn{2}{c}{Attack of \citet{Geiping2020}} \\\cmidrule{2-7}
Dataset & \makecell{Gaussian \\ noise} &\makecell{Gradient \\ compression}  &\makecell{Gaussian \\ noise}  & \makecell{Gradient \\ compression} &\makecell{Gaussian \\ noise} &\makecell{Gradient \\ compression} \\\midrule
EMNIST &0 &0 &0 &0 &0 &0.04 \\
Fashion-MNIST &0 &0 &0 &0 &0 &0.48 \\
CIFAR-10 &0 &0 &0 &0 &0 &0.12 \\
\bottomrule
\end{tabular}
\caption{\dani{Success rate of 25 trials of the reconstruction attacks from  \citep{Zhu2020deep}, \citep{Wei2020} and \citet{Geiping2020} with Gaussian noise and Gradient compression defences.}}\label{tab:rec_defensas}
\end{table}

\begin{figure}%
    \centering
    \includegraphics[trim={0 0 0 0.65cm},clip,width=2.5cm]{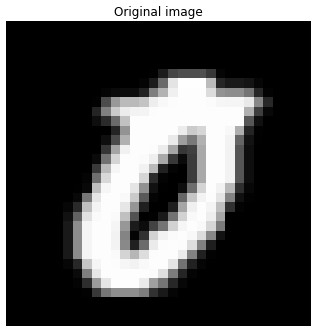} \qquad%
    \includegraphics[trim={0 0 0 0.7cm},clip,width=2.5cm]{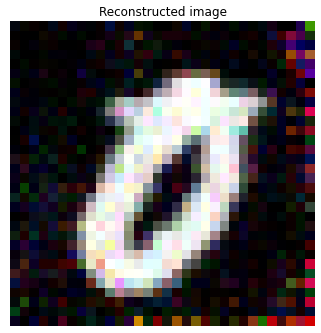}%
    \includegraphics[trim={0 0 0 0.65cm},clip,width=2.5cm]{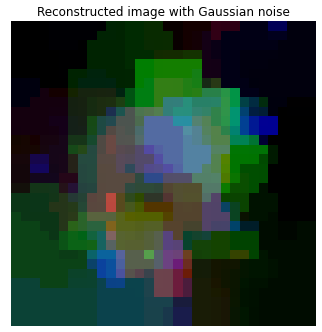}%
    \includegraphics[trim={0 0 0 0.6cm},clip,width=2.5cm]{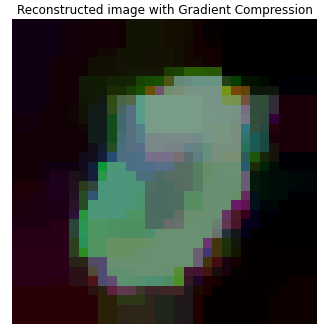}%
    
    \caption{From left to right, reconstruction using the attack of \citet{Geiping2020} of an image with label 0 from EMNIST dataset, without any defence, with Gaussian noise defence and with Gradient compresion defence.}%
    
    \label{fig:mnist-gei}%
\end{figure}

In Table \ref{tab:rec_defensas}, we can observe the stunning performance of both defences as they completely stop the attacks of \citep{Zhu2020deep, Wei2020} from achieving success. While the addition of Gaussian noise of this magnitude is known to reduce the performance of the task \citep{Zhu2020deep}, the gradient compression defence can handle higher compression rates without significantly hurting performance \citep{lin2017deep}. When it comes to the attack of \citet{Geiping2020}, we find that  the Gaussian noise defence is as effective as in the other attacks. This might be due to the differentially private properties of the Gaussian noise. However, the Gradient compression defence fails to completely stop the attack of \citet{Geiping2020}. Specially for the Fashion-MNIST dataset, where almost half of the times the attack succeeded. An example of a reconstruction trial with and without defences is shown in Figure \ref{fig:mnist-gei}, it gives an hint of the behaviour of the attack. Gradient compression is the worst performing defence, probably due to the fact that compressing the gradient does not affect the task of minimizing the cosine similarity between the shared and the reconstructed image gradient.

In conclusion, the Feature inference attacks studied in this section pose a high risk to privacy, as there are many attacks that succeed at extracting private information from gradients. Luckily, there are defences that can thwart the success rate of the attacks and provide privacy without changing significantly the performance of the trained model. However, this is not true for all the analysed attacks, there is still room for improvement as the attack from \citet{Geiping2020} seems to be able to escape them in some situations. Additionally, this threat is not only related to FedSGD scenarios, it is also related to federated scenarios where parameters are exchanged between clients.




\section{Guidelines for the application of defences against adversarial attacks}\label{guidelines}

\dani{Due to the large number of attacks identified, and the wide variety of defences proposed in the literature, it can be difficult to choose which type of defence is appropriate for each situation. Moreover, most defences are designed with the objective of defending against a particular adversarial attack. However, as a collateral benefit, they can prevent the success of other types of adversarial attacks.}

\dani{In this section we provide some guidelines in terms of a summary of which categories of defences work to defend the identified categories of attacks, specifying the degree to which they do so.}

In Table \ref{tab:summary_model} we summarize which categories of defences \dani{are able to defend} against attacks to the model and privacy attacks, respectively. \dani{For the sake of clarity, we represent the relationship between categories of attacks and categories of defences. Hence, when we affirm that a category of defence is able to defend against a category of attacks means that there are at least one defence belonging to that category who is able to defend against them.}

\dani{In this line, we differentiate between:
\begin{itemize}
    \item Complete defence\Hquad\GreenCircle: the defence category is able to stop the attacks of the attack category.
    \item Partial defence\Hquad\OrangeCircle: the defence category is able to significantly reduce the performance of the attacks of the attack category but not stop it.
    \item No defence\Hquad\RedCircle: the performance of the attacks of the attack category is not affected significantly by the defence category.
    \item Unknown defence\Hquad\GrayCircle: there is neither enough experimentation available nor theoretical support to assess the behaviour of the attacks of the attack category with the defence category.
\end{itemize}
}

\begin{table*}[h!]
\begin{center}
\footnotesize
\begin{tabularx}{\textwidth}{XXc@{\hphantom{0}}cc@{\hphantom{0}}c@{\hphantom{0}}c}
\toprule
& & \multicolumn{2}{c}{\textbf{Attacks to the federated model}} & \multicolumn{3}{c}{\textbf{Privacy attacks}}\\ \cmidrule{3-7}
& & \textbf{Untargeted} & \textbf{Targeted} & \textbf{Property} & \textbf{Membership} & \textbf{Feature} \\ \midrule

\multirow{5}{*}{\textbf{Server defences}} & \textbf{Mod. of learning rate} & \GreenCircle & \GreenCircle & \RedCircle & \RedCircle & \RedCircle\\
& \textbf{Robust agg.} & \GreenCircle & \GreenCircle  & \RedCircle & \RedCircle  & \RedCircle\\
& \textbf{Anomaly detection} & \GreenCircle &  \GreenCircle  & \RedCircle & \RedCircle  & \RedCircle\\
& \textbf{Based on DP} & \OrangeCircle & \OrangeCircle & \OrangeCircle & \GreenCircle & \GreenCircle\\
& \textbf{Less is more} & \GreenCircle & \OrangeCircle & \RedCircle & \RedCircle & \RedCircle \\ \midrule

\multirow{3}{*}{\textbf{Client defences}} & \textbf{Based on DP} & \OrangeCircle & \OrangeCircle  & \OrangeCircle & \GreenCircle & \GreenCircle\\
& \textbf{Optimized training} & \GreenCircle & \OrangeCircle & \GrayCircle & \GrayCircle & \GrayCircle\\
& \textbf{Perturbations methods} & \GrayCircle & \GrayCircle  & \GrayCircle & \GreenCircle & \GreenCircle\\ \midrule

\multirow{2}{*}{\textbf{Comm. channel}} & \textbf{Blockchain} & \OrangeCircle & \OrangeCircle & \OrangeCircle & \OrangeCircle & \OrangeCircle \\
& \textbf{SMPC} & \OrangeCircle & \OrangeCircle & \OrangeCircle & \OrangeCircle & \OrangeCircle  \\
\bottomrule
\end{tabularx}
\end{center}
\caption{Summary of application of the defences to adversarial attacks, both attacks to the federated model and privacy attacks.}\label{tab:summary_model}
\end{table*}

The summary of the state of the art provided in the Table \ref{tab:summary_model} allows us to draw the following conclusions:
\begin{enumerate}
    \item In general, defences \dani{based on DP, which are} designed to defend against privacy attacks, partially defend against attacks to the model\dani{, specially those based on DP,} but not the other way around. 
    \item Broadly speaking, the defence against attacks to the model is more settled than the defences against privacy attacks. In particular, for property inference attacks, there is no defence considered as complete.
    \item There is still a long way to go in designing defences to prevent attacks in FL and, crucially, to find a defence that prevents from all types of attacks at the same time.
\end{enumerate}

\section{Lessons learned}\label{sec:lessons}

Based on the extensive research and analysis of the available works, we have built up the taxonomy proposed in this paper. However, what has been learned goes beyond this. To sum up, the lessons learned are: 

\begin{enumerate}
    \item The \dani{identification} of \eugenio{vulnerabilities in the form of adversarial attacks} and \dani{the proposal of} defences against them in FL is a field of research in continuous development. The volume to date of scientific work covering these challenges is growing and is not likely to diminish in the coming years.
    
    \item \dani{Attacks to the federated model are easier to defend against than the privacy-attacks. However, they have shown much greater effectiveness, with even the simplest attacks being detrimental to the model.}
    
    \item Privacy attacks require very peculiar settings to achieve a reasonable success, that is, most of the assumptions required to perform them are very hard to achieve in real FL scenarios. Such scenarios are \dani{usually} bounded by the lack of the following resources: data, raw computing power and time.
    
    \item Most defences against inference attacks, although designed for inference attacks, dissipate the performance of attacks against the \dani{federated} model, but not the other way around. Therefore, the use of DP-inspired mechanisms will be crucial if we want to defend against generic category attacks.
    
    \item The implementation of \dani{defences} based on DP and based on perturbation methods require extensive \dani{fine-tuning} in order to provide a nice trade-off between privacy and performance. Most of the \dani{defences} require access to big computational resources or they will be too slow to apply. Therefore, such defences might not be suitable for FL settings with low power devices. Additionally, to our knowledge, there is not a \dani{consensus} about how to measure the trade-off between privacy and performance.
\end{enumerate}

\paco{To finish, as a fundamental lesson learned is} that the field of \dani{adversarial} attacks and defences in FL is a research area in steady development, which is not expected to change in the forthcoming years. There are still many vulnerabilities which need to be faced in order to ensure a truly secure and privacy-preserving learning environment.

\section{\new{Challenges of addressing federated learning threats}}\label{sec:challenges}

\eugenio{Regarding the previous lessons learnt, we identified the following challenges that the field will have to face up in the next years.}

\paragraph{\textbf{Defences vs. attacks}} An identified trend is that for each defence proposed, it is possible to \dani{identify a vulnerability that can be turned into an adversarial attack}, and vice versa. Therefore, one of the biggest challenges is to find both: (1) \dani{all vulnerabilities present in a FL scenario that an attacker could exploit}, and (2) a defence effective enough to defend against any attack. For the time being, this seems a long way off, as the different perspectives from which both problems have been approached are ad hoc to the type of attack to be \dani{identified} or defended. From our point of view, the study of defences is \dani{crucial}, since the final goal is to achieve a secure\eugenio{, robust} and private learning environment. Along this vein, the optimal defence will be the one that combines the best proposals in each of the categories, in such a way that it manages to defend against all types of attacks while maintaining performance in the original task. There are existing approaches that combine client's filtering with noise addition \cite{ozdayi2020defending}, although there is still a long way to go.

\paragraph{\textbf{Trade-off in defences}} In most defences, we find that it is difficult to strike a trade-off between preventing the model from attacking and not impairing performance in the original task. For example, in those based on DP, we find that in order to ensure data privacy, a large amount of noise has to be added, which significantly impairs the performance of the model \cite{bagdasaryan2019differential}. \dani{Therefore one of the main challenges would be the development of more efficient DP methods, and the extension of DP to defences against all adversarial attacks.} This situation also occurs in defences based on client filtering when more clients than necessary are filtered out, thus losing information in the aggregation process. In short, it is difficult to strike a trade-off between preventing an attack and not losing or poisoning the information received by clients.
    
\paragraph{\textbf{Non-IID assumptions}} The non-IID nature of the training data distributed among clients often makes it difficult to differentiate between adversarial clients and those who have had a very \dani{different from the rest}, but still valuable, learning process.  One \dani{common approach} is to use anomaly detection algorithms suitable for non-IID distributions \cite{pang2021homophily} or approaches which not rely on data distribution \cite{DBLP:journals/corr/abs-2011-01767}, however, there are still problems in differentiating between clients with a highly skewed distribution and adversarial clients in most cases.

\paragraph{\textbf{Generalised FL}} The vast majority of adversarial attacks \dani{have been identified} for HFL. In particular, the adversarial attacks to the federated model. Although there is already existing work on privacy attacks in VFL \cite{Li2021}, \dani{there is still a long way to go in identifying and analysing the vulnerabilities in terms of leakage of information of attacker possibilities in other categories of FL wich are becoming widely used such as VFL or FTL \cite{10.1145/3298981}}. Therefore, we believe that in the coming years, work on \dani{identifying} attacks for VFL and FTL \dani{and the research in defences against them} will take centre stage.
    
\paragraph{\textbf{Combination with other trends}} While ensuring data privacy is one of the main goals of FL, there are other desirable features. For example, some of the most popular trends are Personalised FL \cite{tan2021personalized} or fairness in FL \cite{ezzeldin2021fairfed}. We believe that, at the end of the day, data security and privacy must be a requirement in all other approaches. Therefore, several future works will address this issue as a cross-cutting objective while developing proposals for more concrete desirable features. For example, a method of personalised FL that is secure against adversarial attacks.

\section{\paco{Conclusions}}\label{sec:conclusions}

\paco{FL emerges as a solution to the computational costs and privacy-preserving demands of the most groundbreaking ML. However, this new learning paradigm brings new challenges, particularly in terms of adversarial attacks and defending against them. Hence, several proposals of new adversarial attacks or adaptations of centralised ones as well as defences ad hoc to these attacks have been proposed in the recent years.}

\paco{We proposed several taxonomies according to different criteria that eases the knowledge of the wide field of FL threats. In addition, we conducted an extensive experimental study which leads us to propose a guidelines for the application of defences against adversarial attacks, and to highlight a set of lessons learned and challenges related to FL threats.}



\new{We conclude that the study of FL threats is an ongoing field of research, due to its importance in ensuring FL as a robust machine learning paradigm that safeguards data privacy. There are still several challenges to be faced and directions to be studied in order to identify additional threats (or vulnerabilities) of FL, as well as the appropriate mechanisms to make it a resilient and robust learning paradigm against those threats.}









%
\section*{Acknowledgments} \label{sec:acknowledgment}

This research work is partially supported by the Trust-ReDaS (PID2020-119478GB-I00), the FedDAP (PID2020-116118GA-I00) and the EQC2018-005084-P projects from the Spanish Government, and a grant from the European Regional Development Fund (ERDF). Nuria Rodríguez Barroso and Eugenio Martínez Cámara were supported by the Spanish Government fellowship programmes Formación de Profesorado Universitario (FPU18/04475) and Juan de la Cierva Incorporación (IJC2018-036092-I) respectively.

\newpage

\bibliography{main}

\bibliographystyle{elsarticle-num-names} 

\end{document}

%% file: main_categorisation.tex
\begin{center}
\resizebox{0.5\textwidth}{!}{\begin{tikzpicture}[mindmap,
  level 1 concept/.append style={level distance=130,sibling angle=30},
  extra concept/.append style={color=blue!50,text=black}]

    \definecolor{rojo_pastel}{HTML}{FFABAB}
    \definecolor{azul_pastel}{HTML}{6EB5FF}
    \definecolor{verde_pastel}{HTML}{ADCFA3}

 
  \begin{scope}[mindmap, concept color=rojo_pastel, text=white]
    \node [concept] (root) at (0, 0) {\textbf{Adversarial attacks in FL}};
  \end{scope}


  \begin{scope}[mindmap, concept color=azul_pastel, text=white]
    \node [concept] (property) at (-5, -5) {Attacks to the federated model};
  \end{scope}

  \begin{scope}[mindmap, concept color=verde_pastel, text=white]
    \node [concept] (feature) at (5, -5) {Privacy attacks};
  \end{scope}

    \path (root) to[circle connection bar switch color=from (rojo_pastel) to (azul_pastel)] (property);
    \path (root) to[circle connection bar switch color=from (rojo_pastel) to (verde_pastel)] (feature);

\end{tikzpicture}}
\end{center}

%% file: attacks_to_the_model.tex
\usetikzlibrary{mindmap,backgrounds,shapes.misc}

\begin{center}
\resizebox{0.9\textwidth}{!}{
\begin{tikzpicture}[mindmap,
  level 1 concept/.append style={level distance=130,sibling angle=30},
  extra concept/.append style={color=morado_pastel!50,text=black}]

    \definecolor{rojo_pastel}{HTML}{FFABAB}
    \definecolor{azul_pastel}{HTML}{6EB5FF}
    \definecolor{verde_pastel}{HTML}{ADCFA3}
    \definecolor{morado_pastel}{HTML}{CAA7BD}


  \begin{scope}[mindmap, concept color=rojo_pastel, text=white]
    \node [concept] at (-4, 3) {According to the attack moment}
      child [grow=30, level distance=120] {node [concept] (test) {Inference-time attacks}}
      child [grow=-20, level distance=120] {node [concept] (training) {Training-time attacks}};
  \end{scope}


  \begin{scope}[mindmap, concept color=azul_pastel,text=white]
    \node [concept] (frequency) at (-8,-2) {According to the frequency}
      child [grow=-50, level distance=120]
        {node [concept] (mul) {Multiple}}
      child [grow=0, level distance=120] 
        {node [concept] (one) {One shot}};
  \end{scope}

    \begin{scope}[mindmap, concept color=morado_pastel,text=white]
    \node [concept] (objective) at (0,-12) {According to the objective}
      child [grow=120, level distance=120]
        {node [concept] (tar) {Targeted}}
      child [grow=60, level distance=120] 
        {node [concept] (untar) {Untargeted}};
  \end{scope}

    \begin{scope}[mindmap, concept color=verde_pastel,text=white]
    \node [concept] (part) at (8,-2) {According to the poisoned part}
      child [grow=220, level distance=120]
        {node [concept] (data) {Data-poisoning}}
      child [grow=180, level distance=120] 
        {node [concept] (model) {Model-poisoning}};
  \end{scope}


    \path (one) to[circle connection bar switch color=from (black!15) to (black!15)] (tar);
    \path (one) to[circle connection bar switch color=from (black!15) to (black!15)] (untar);
    \path (mul) to[circle connection bar switch color=from (black!15) to (black!15)] (tar);
    \path (mul) to[circle connection bar switch color=from (black!15) to (black!15)] (untar);
    \path (data) to[circle connection bar switch color=from (black!15) to (black!15)] (tar);
    \path (data) to[circle connection bar switch color=from (black!15) to (black!15)] (untar);
    \path (model) to[circle connection bar switch color=from (black!15) to (black!15)] (tar);
    \path (model) to[circle connection bar switch color=from (black!15) to (black!15)] (untar);

    \path (training) to[circle connection bar switch color=from (rojo_pastel) to (azul_pastel)] (frequency);
    \path (training) to[circle connection bar switch color=from (rojo_pastel) to (morado_pastel)] (objective);
    \path (training) to[circle connection bar switch color=from (rojo_pastel) to (verde_pastel)] (part);
    
     \node[rectangle, draw, fill=black!15, fill opacity=0.3,draw = lightgray, rounded corners = 10pt, text opacity = 1, text = gray, text width=4cm, minimum height = 2cm] (r) at (0,-2.5) {\large \textbf{Taxonomy of attacks to the federated model}};

\end{tikzpicture}}
\end{center}

%% file: backdoor_attacks.tex
\begin{center}
\resizebox{0.8\textwidth}{!}{\begin{tikzpicture}[mindmap,
  level 1 concept/.append style={level distance=130,sibling angle=30},
  extra concept/.append style={color=verde_pastel!50,text=black}]

    \definecolor{rojo_pastel}{HTML}{FFABAB}
    \definecolor{azul_pastel}{HTML}{6EB5FF}
    \definecolor{verde_pastel}{HTML}{ADCFA3}
    \definecolor{morado_pastel}{HTML}{CAA7BD}

  \begin{scope}[mindmap, concept color=rojo_pastel, text=white]
    \node [concept] at (-20, 3) {\textbf{Targeted/ Backdoor attacks}}
      child [grow=30, level distance=120] {node [concept] (input) {Input-instance-key}
      }
      child [grow=-20, level distance=120] {node [concept] (pattern) {Pattern-key}};
  \end{scope}


  \begin{scope}[mindmap, concept color=azul_pastel,text=white]
    \node [concept, scale=0.8] (design) at (-12,7.5) {Design of the pattern}
      child [grow=-20, level distance=120]
        {node [concept] (blended) {Blended Injection}
        child [grow=25, level distance=80] {node [concept] (mix) {Blended-Accessory Injection}}
        }
      child [grow=20, level distance=120] 
        {node [concept] (accessory) {Accessory Injection}};
  \end{scope}

    \begin{scope}[mindmap, concept color=verde_pastel, text=white]
    \node [concept, scale=0.8] (number) at (-12, 2) {Number of patterns}
      child [grow=20, level distance=120] {node [concept] (single) {Single}
      }
      child [grow=-20, level distance=120] {node [concept] (multiple) {Multiple}};
  \end{scope}
  
      \begin{scope}[mindmap, concept color=morado_pastel, text=white]
    \node [concept, scale=0.8] (var) at (-12, -3.5) {Variability}
      child [grow=20, level distance=120] {node [concept] (static) {Static}
      }
      child [grow=-20, level distance=120] {node [concept] (dyn) {Dynamic}};
  \end{scope}

    \path (pattern) to[circle connection bar switch color=from (rojo_pastel) to (azul_pastel)] (design);
    \path (mix) to[circle connection bar switch color=from (azul_pastel) to (azul_pastel)] (accessory);
    \path (pattern) to[circle connection bar switch color=from (rojo_pastel) to (verde_pastel)] (number);
    \path (pattern) to[circle connection bar switch color=from (rojo_pastel) to (morado_pastel)] (var);
\end{tikzpicture}}
    
\end{center}

%% file: poisoned.tex
\begin{center}
\resizebox{0.8\textwidth}{!}{\begin{tikzpicture}[mindmap,
  level 1 concept/.append style={level distance=130,sibling angle=30},
  extra concept/.append style={color=blue!50,text=black}]

    \definecolor{rojo_pastel}{HTML}{FFABAB}
    \definecolor{azul_pastel}{HTML}{6EB5FF}
    \definecolor{verde_pastel}{HTML}{ADCFA3}
    \definecolor{morado_pastel}{HTML}{CAA7BD}

  \begin{scope}[mindmap, concept color=rojo_pastel, text=white]
    \node [concept] (root) at (0, 0) {\textbf{Poisoning attacks}};
  \end{scope}


  \begin{scope}[mindmap, concept color=azul_pastel,text=white]
    \node [concept] (property) at (-5, -5) {Data-poisoning attacks}
        child [grow=135, level distance=120] {node [concept] (ind) {Label-flipping}}
        child [grow=180, level distance=120] {node [concept] (population) {Poisoning samples}}
        child [grow=225, level distance=120] {node [concept] (out-of) {Out-of-distribution}};
  \end{scope}

  \begin{scope}[mindmap, concept color=verde_pastel,text=white]
    \node [concept] (feature) at (5, -5) {Model-poisoning attacks}
        child [grow=45, level distance=120] {node [concept] (ind) {Random weights}}
        child [grow=0, level distance=120] {node [concept] (population) {Optimisation methods}}
        child [grow=-45, level distance=120] {node [concept] (out-of) {Information Leakage}};
  \end{scope}

    \path (root) to[circle connection bar switch color=from (rojo_pastel) to (azul_pastel)] (property);
    \path (root) to[circle connection bar switch color=from (rojo_pastel) to (verde_pastel)] (feature);

\end{tikzpicture}}
\end{center}

%% file: privacy_attacks.tex
\begin{center}
\resizebox{0.8\textwidth}{!}{\begin{tikzpicture}[mindmap,
  level 1 concept/.append style={level distance=130,sibling angle=30},
  extra concept/.append style={color=verde_pastel!50,text=black}]

    \definecolor{rojo_pastel}{HTML}{FFABAB}
    \definecolor{azul_pastel}{HTML}{6EB5FF}
    \definecolor{verde_pastel}{HTML}{ADCFA3}
    \definecolor{morado_pastel}{HTML}{CAA7BD}

  \begin{scope}[mindmap, concept color=rojo_pastel, text=white]
    \node [concept] (root) at (0, 0) {\textbf{Privacy attacks}};
  \end{scope}


  \begin{scope}[mindmap, concept color=azul_pastel,text=white]
    \node [concept] (property) at (6, 5) {Property Inference}
        child [grow=-30, level distance=120] {node [concept] (ind) {Individual distribution}}
        child [grow=30, level distance=120] {node [concept] (population) {Population distribution}};
  \end{scope}
  
  \begin{scope}[mindmap, concept color=verde_pastel,text=white]
    \node [concept] (member) at (6, 0) {Membership Inference};
  \end{scope}
  
  \begin{scope}[mindmap, concept color=morado_pastel,text=white]
    \node [concept] (feature) at (6, -5) {Feature Inference};

  \end{scope}
  
  \begin{scope}[mindmap, concept color=morado_pastel,text=white]
    \node [concept, scale=0.8] (rec) at (12, -5) {Reconstruction}
        child [grow=30, level distance=120] {node [concept] (param) {Parameter based}}
        child [grow=-30, level distance=120] {node [concept] (grad) {Gradient based}};
  \end{scope}
  
    \path (root) to[circle connection bar switch color=from (rojo_pastel) to (azul_pastel)] (property);
    \path (root) to[circle connection bar switch color=from (rojo_pastel) to (verde_pastel)] (member);
    \path (root) to[circle connection bar switch color=from (rojo_pastel) to (morado_pastel)] (feature);
    \path (member) to[circle connection bar switch color=from (verde_pastel) to (morado_pastel)] (feature);
    \path (rec) to[circle connection bar switch color=from (morado_pastel) to (morado_pastel)] (feature);

\end{tikzpicture}}
\end{center}

%% file: defences.tex
\usetikzlibrary{mindmap,backgrounds}

\begin{center}
\resizebox{0.8\textwidth}{!}{
\begin{tikzpicture}[mindmap,
  level 1 concept/.append style={level distance=130,sibling angle=30},
  extra concept/.append style={color=verde_pastel!50,text=black}]

    \definecolor{rojo_pastel}{HTML}{FFABAB}
    \definecolor{azul_pastel}{HTML}{6EB5FF}
    \definecolor{verde_pastel}{HTML}{ADCFA3}
    \definecolor{morado_pastel}{HTML}{CAA7BD}

  \begin{scope}[mindmap, concept color=rojo_pastel, text=white]
    \node [concept] (def) at (0, 0) {\textbf{Defences}};
  \end{scope}
  
    \begin{scope}[mindmap, concept color=azul_pastel, text=white]
    \node [concept] (ser) at (0, 5.5) {\textbf{Server}}
        child [grow=-30, level distance=120] {node [concept] {Robust aggregation}}
        child [grow=30, level distance=120] {node [concept] {Anomaly detection}}
        child [grow=90, level distance=120] {node [concept] {Based on DP}}
        child [grow=150, level distance=120] {node [concept] {Mod. of learning rate}}
        child [grow=210, level distance=120] {node [concept] {Less is more}};
  \end{scope}
  
    \begin{scope}[mindmap, concept color=verde_pastel, text=white]
    \node [concept] (cli) at (-4, -4) {\textbf{Communication channel}}
        child [grow=180, level distance=120] {node [concept] {Blockchain}
      }
      child [grow=240, level distance=120] {node [concept] {SMPC}};
  \end{scope}
  
    \begin{scope}[mindmap, concept color=morado_pastel, text=white]
    \node [concept] (com) at (4, -4) {\textbf{Client}}
      child [grow=0, level distance=120] {node [concept] {Based on DP}}
      child [grow=-120, level distance=120] {node [concept] {Perturbation methods}}
      child [grow=-60, level distance=120] {node [concept] {Optimized training}};
  \end{scope}
  
    \path (def) to[circle connection bar switch color=from (rojo_pastel) to (azul_pastel)] (ser);
    \path (def) to[circle connection bar switch color=from (rojo_pastel) to (verde_pastel)] (cli);
    \path (def) to[circle connection bar switch color=from (rojo_pastel) to (morado_pastel)] (com);

\end{tikzpicture}}
\end{center}